\documentclass[fleqn,12pt]{wlscirep}
\usepackage[utf8]{inputenc}
\usepackage[T1]{fontenc}

\usepackage{braket}
\newcommand{\bvec}[1]{\boldsymbol #1}
\addto\captionsenglish{\renewcommand{\figurename}{Fig.}}

\title{Unconventional Non-local Relaxation Dynamics in a Twisted Graphene Moir\'e Superlattice}

\author[1,*,$\dagger$]{Dorri Halbertal}
\author[1,2,$\dagger$]{Simon Turkel}
\author[3]{Christopher J. Ciccarino}
\author[4]{Jonas Profe}
\author[1]{Nathan Finney}
\author[1]{Valerie Hsieh}
\author[5]{Kenji Watanabe}
\author[6]{Takashi Taniguchi}
\author[1]{James Hone}
\author[1]{Cory Dean}
\author[3]{Prineha Narang}
\author[1,2]{Abhay N. Pasupathy}
\author[4,7]{Dante M. Kennes}
\author[1]{D. N. Basov}
\affil[1]{Department of Physics, Columbia University, New York, NY 10027, USA.}
\affil[2]{Condensed Matter Physics and Materials Science Division, Brookhaven National Laboratory, Upton, NY 11973, USA.}
\affil[3]{Harvard John A. Paulson School of Engineering and Applied Sciences, Harvard University, Cambridge, MA 02138, USA.}
\affil[4]{Institute for Theory of Statistical Physics, RWTH Aachen University, and JARA Fundamentals of Future Information Technology, 52062 Aachen, Germany.}
\affil[5]{Research Center for Functional Materials, National Institute for Materials Science, 1-1 Namiki, Tsukuba 305-0044, Japan.}
\affil[6]{International Center for Materials Nanoarchitectonics, National Institute for Materials Science, 1-1 Namiki Tsukuba 305-0044, Japan.}
\affil[7]{Max Planck Institute for the Structure and Dynamics of Matter, Center for Free Electron Laser Science, Hamburg, Germany.}

\affil[*]{dorrihal@gmail.com}
\affil[$\dagger$]{These authors contributed equally.}

\begin{abstract}
 The electronic and structural properties of atomically thin materials can be controllably tuned by assembling them with an interlayer twist.  During this process, constituent layers spontaneously rearrange themselves in search of a lowest energy configuration.  Such relaxation phenomena can lead to unexpected and novel material properties.  Here, we study twisted double trilayer graphene (TDTG) using nano-optical and tunneling spectroscopy tools. We reveal a surprising optical and electronic contrast, as well as a stacking energy imbalance emerging between the moir\'e domains.  We attribute this contrast to an unconventional form of lattice relaxation in which an entire graphene layer spontaneously shifts position during fabrication.  We analyze the energetics of this transition and demonstrate that it is the result of a non-local relaxation process, in which an energy gain in one domain of the moir\'e lattice is paid for by a relaxation that occurs in the other.  

\end{abstract}

\begin{document}

\flushbottom
\maketitle
\thispagestyle{empty}


\textbf{Main Text:} The discovery of superconductivity in rotationally misaligned graphene bilayers established moir\'e engineering as a robust way to create strongly correlated phases in van der Waals heterostructures \cite{Cao2018_insulator,Cao2018_superconductor}. Since this initial discovery, a wide range of moir\'e materials have emerged with fascinating electronic properties such as correlated insulators\cite{Xie2021,Nuckolls2020,Liu2020,Cao2020_tdbg,He2021,PhysRevLett.123.197702}, strange metals \cite{Ghiotto2021,Jaoui2022}, electronic nematics\cite{Rubio-Verd2022,Samajdar_2021,Kerelsky2019,Jiang2019}, and Wigner crystals\cite{Li2021}, among other unconventional phases \cite{Kennes2021}.  With the advent of new and more complex moir\'e device geometries, with greater than two layers\cite{Cao2020_tdbg,PhysRevLett.123.197702,He2021,Liu2020} or greater than one twist angle between layers\cite{doi:10.1126/science.abg0399,Park2021,2112.07127,2112.07841}, it becomes increasingly important to consider the effects of lattice relaxation, or the spontaneous rearrangement of atoms in search of a lower energy configuration, on the final microscopic crystal structure.  In mirror symmetric twisted trilayer graphene, for instance, it was recently observed\cite{Turkel2022} that lattice relaxation leads to the emergence of moir\'e defects that are not observed in the simpler twisted bilayer system.  A fuller understanding of relaxation phenomena in moir\'e heterostructures thus holds the potential to enable exploitation of these effects as a means of engineering novel or otherwise unstable material systems.\\

Twisted double trilayer graphene (TDTG), a moir\'e material that has not yet been experimentally investigated, is a natural next step in extending the moir\'e paradigm to more complex structures, in which lattice relaxation can lead to unexpected atomic configurations. TDTG is formed in a manner analogous to twisted bilayer (TBG) and twisted double bilayer graphene (TDBG), namely by stacking two Bernal trilayers from the same source crystal with a small relative twist.  In the low twist angle limit, moir\'e patterns in few-layer graphene (FLG) systems generate large domain structures with distinct crystallographic stackings\cite{Halbertal2021, Kerelsky2021}.  If rigidly stacked in a manner that preserves the ABA stacking of each trilayer (referred to here as the ``rigid'' scenario), TDTG forms domains of mixed rhombohedral and Bernal character with ABABCB and BCBABA stackings (Fig. \ref{fig:TDTG_overview}a), where each domain contains a unit of three rhombohedrally stacked layers (3R). \\

It is conceivable, however, that stresses and strains applied to a sample during the fabrication process can act as an effective annealing, allowing the system to explore other stacking configurations before reaching the lowest energy equilibrium.  In the case of TDTG, applying a simple translation to the second layer (Fig. \ref{fig:TDTG_overview}b) results in one domain (ABABAB) with pure Bernal stacking and another domain (BCBACA) with a unit of four rhombohedrally stacked layers (4R).  This ``layer slide'' scenario provides an interesting test case from an energy standpoint.  While the pure Bernal phase is energetically favorable, its realization comes at the price of increasing the local stacking energy of the mixed rhombohedral domain by transforming it from a 3R to a 4R configuration.  Prior studies of relaxation effects in twisted van der Waals heterostructures\cite{Halbertal2021,PhysRevB.96.075311,Haddadi2020,PhysRevB.99.205134,moore_nanoscale_2021} have focused on processes in which lattice relaxation acts to uniformly reduce the stacking energy at every location in space, generally by minimizing the area of higher energy domains and maximizing the area of their lower energy counterparts, as occurs in minimally twisted TDBG.  Consideration of the scenarios presented in Fig. \ref{fig:TDTG_overview}a,b for TDTG raises the question of whether relaxation processes can likewise act to offset an increase of the stacking energy in one area of a device with a decrease in an area that is as far as several microns away.  Such relaxation at a distance would offer a means of stabilizing elusive atomic configurations with distinct electronic properties by structurally coupling them to low energy phases through moir\'e patterning.\\

In this work, we utilize mid-infrared scanning near-field optical microscopy (SNOM) and scanning tunneling microscopy (STM) and spectroscopy (STS) to characterize the optical and electronic properties of TDTG samples.  The use of SNOM and STM/S on identical samples allows us to characterize the electronic properties of a device over both large (microns) and small (nanometers) areas, giving direct experimental access to the interplay of length scales that is crucial to non-local relaxation dynamics.  Fig. \ref{fig:TDTG_overview}c,d shows phase resolved SNOM images\cite{Sunku2018h,Hesp2021,Jiang2016a} taken over a $\sim10~\mu\textrm{m}$ region of the TDTG device shown in the inset of Fig. \ref{fig:TDTG_overview}c.  In the rigid scenario for TDTG (Fig. \ref{fig:TDTG_overview}a), we expect the local stacking configuration in each domain to be ABABCB and BCBABA, which are related to each other by inversion across the $x$-$y$ plane ($C_2^z$).  In the absence of external $C_2^z$ symmetry breaking mechanisms, such as displacement field or an asymmetric dielectric environment, these two configurations are therefore expected to possess identical electronic and structural properties.  Contrary to this expectation, a moir\'e superlattice with clear optical contrast between adjacent domains is observed both in the amplitude (Fig. \ref{fig:TDTG_overview}c) and phase (Fig. \ref{fig:TDTG_overview}d) of the near-field signal.  Furthermore, there is a clear imbalance in the stacking energy of the two domains, as evidenced by the convexity (concavity) of the bright (dark) triangles in Fig. \ref{fig:TDTG_overview}d.  This difference in stacking energy is particularly clear in regions of large heterostrain, such as in the top left of Fig. \ref{fig:TDTG_overview}d, where a linear pattern is observed, corresponding to double domain walls (DDWs) that emerge due to the collapse of unstable domains \cite{Halbertal2021}.\\

Fig. \ref{fig:TDTG_tomography_experiment} examines the TDTG moir\'e superlattice at the sub-micron length scale.  The complex near-field signal is presented in Fig. \ref{fig:TDTG_tomography_experiment}a-d as a function of tapping-probe demodulation harmonic (focusing on the green square in Fig. \ref{fig:TDTG_overview}c-d). These finer scans further demonstrate the optical contrast between adjacent domains as well as the curving of the domain walls (DWs). Additional details emerge in these higher resolution images as well, including clear bright features along the DWs and bright spots at the DW intersections in Fig. \ref{fig:TDTG_tomography_experiment}d. Furthermore, we find that the complex near-field optical contrast between the domains is a monotonically increasing function of demodulation harmonic, as summarized in Fig. \ref{fig:TDTG_tomography_experiment}e.  We examine the microscopic electronic structure in greater detail by performing STM/S measurements over regions of the same TDTG sample.  The local density of states (LDOS) measured near the Fermi level is displayed in Fig. \ref{fig:TDTG_tomography_experiment}f, which shows a striking contrast in tunneling conductivity between different domains of the moir\'e lattice.  Characteristic spectra acquired on each of the two domains are plotted in Fig. \ref{fig:TDTG_tomography_experiment}g, demonstrating that the observed LDOS contrast is caused by a large spectral peak near zero energy that is present in only the convex domain.  The spectral shape in the concave domain, on the other hand, is largely featureless at low energy and possesses no comparable peak.  Moreover, measurements of the tunneling spectrum on the untwisted region (Fig. \ref{fig:TDTG_tomography_experiment}h) indicate that the source crystal is Bernal trilayer, confirming that the moir\'e contrast is a property of the six-layer system.  This suite of measurements unambiguously demonstrates a significant difference in electronic structure between each of the observed moir\'e superlattice domains.  In the Supplementary Information (section \ref{nearfield_tomography}) we consider and rule out alternative sources of $C_2^z$ symmetry breaking, including the effect of the hBN substrate and the possibility of atomic-scale near-field tomography, i.e. the breaking of $C_2^z$ symmetry by the sharp probe interacting with individual atomic layers.  The naive expectation of the rigid scenario (Fig. \ref{fig:TDTG_overview}a) is therefore clearly not realized in our TDTG device. \\

Applying a global translation to one of the six layers in a TDTG heterostructure has the potential to lower the overall stacking energy of a device even as it might raise the energy density in certain confined regions.  Once we accept that the energy barrier to such a transition can be overcome (see below), there is in principle no reason to restrict our analysis to the particular layer slide scenario depicted in Fig. \ref{fig:TDTG_overview}b.  In an effort to match the experimental observations, we have therefore performed DFT calculations of the band structure, density of states, and stacking energy density of all $2^5$ possible TDTG stacking configurations (each of the five layer interfaces can be stacked as either AB or BA).  Eight of these configurations describe, in the minimally twisted regime, moir\'e pairs that are related by $C_2^z$ symmetry (see Supplementary Information section \ref{c2_pairs}), which we have already ruled out above.  \\

\begin{figure} [htbp]
\centering
\includegraphics[width=\linewidth]{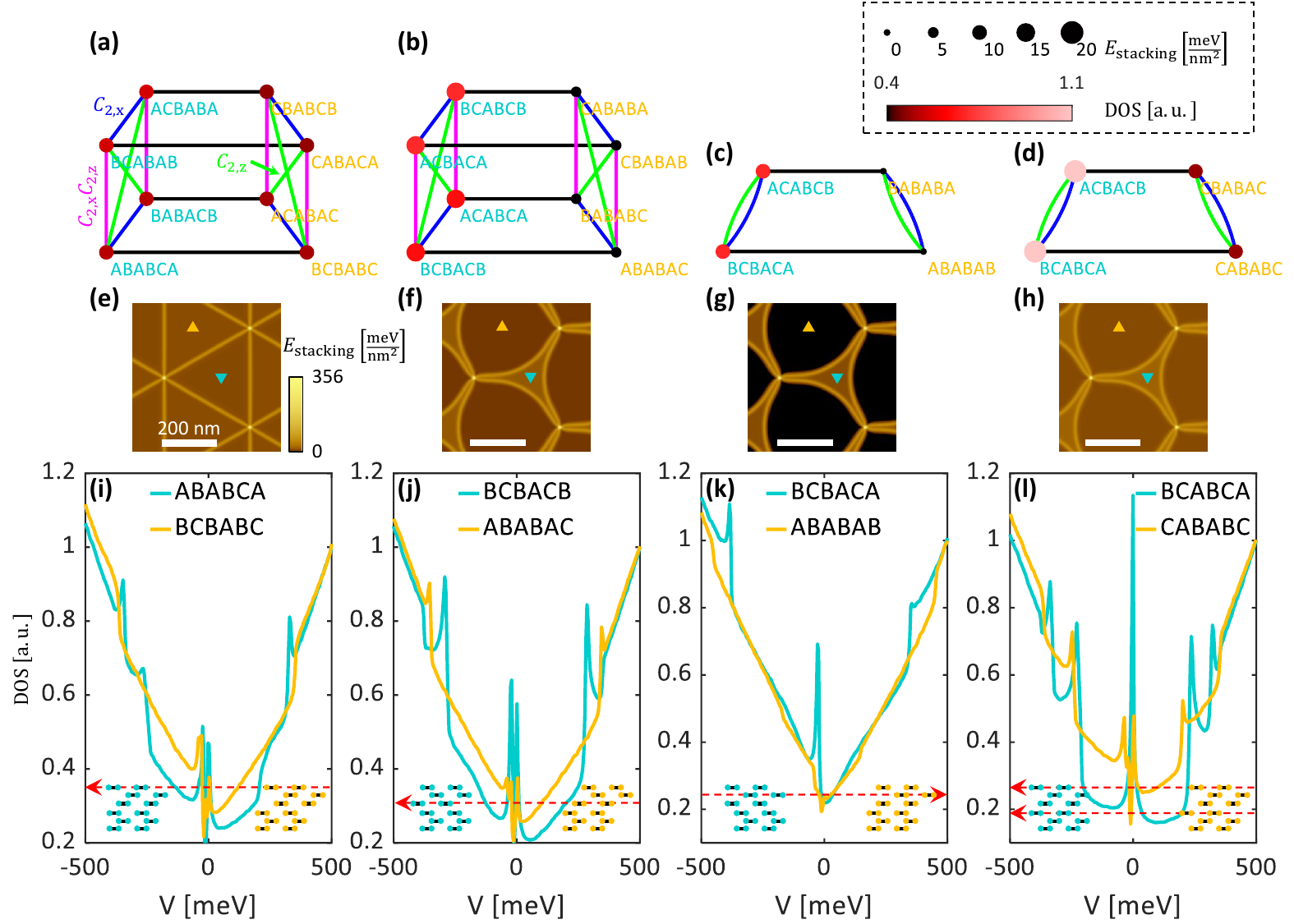}
\caption{\textbf{Exploration of candidates for TDTG moir\'e superlattice structures}. \textbf{(a-d)} Each panel addresses a group of configurations. Each moir\'e pair is connected by a horizontal black line. Configurations that are $C_2$ symmetry pairs are connected by blue ($C_2^x$), green ($C_2^z$) and magenta ($C_2^xC_2^z$) lines respectively. Each configuration is marked by a circle whose color indicates the low energy spectral weight, and whose size indicates the configuration's stacking energy density (see legend).  \textbf{(e-h)} The domain formation is reflected by the stacking energy density for a twist angle of $0.04^\circ$ for each moir\'e pair. Each panel is the result of the atomic relaxation calculation (see Methods) using the DFT calculated energy imbalance of the corresponding moir\'e pair (triangles indicate the phases with matching colors to the configuration text in \textbf{(a-d)}). \textbf{(i-l)} DFT calculated electronic densities of states for different moir\'e pairs (see Methods). The inset shows the lattice structure for each configuration (with consistent colors as in \textbf{(a-d)}). The red arrow indicates the required global layer sliding in order to realize the particular moir\'e superlattice from the rigid ABABCB/BCBABA configuration. The configurations where the moir\'e pairs are also $C_2$ symmetry pairs were omitted here for brevity, as they could not produce an energy imbalance (these missing configurations are explored in Supplementary Information section \ref{c2_pairs}).}
\label{fig:TDTG_all_configurations}
\end{figure}

Fig. \ref{fig:TDTG_all_configurations} explores the remaining twenty-four TDTG configurations with crystallographically distinct moir\'e domains. These can be divided into four groups (corresponding to the four columns of Fig. \ref{fig:TDTG_all_configurations}) based on their symmetries. Moir\'e pairs in Fig. \ref{fig:TDTG_all_configurations}a-d are connected by black horizontal lines ($C_2$ symmetry pairs are connected by colored lines as indicated).  Each possible moir\'e domain is characterized by its calculated Fermi level spectral weight and stacking energy density (reflected in Fig. \ref{fig:TDTG_all_configurations}a-d by vertex color and size respectively).  Moir\'e pairs with a large difference in stacking energy result in curved domains, similar to those seen experimentally, as confirmed by atomic relaxation calculations (Fig. \ref{fig:TDTG_all_configurations}e-h).  In seeking a match with our experimental observations, we therefore require a pair of moir\'e domains with a large difference in both stacking energy and Fermi level spectral weight to match the domain curvature and electronic contrast revealed by SNOM and STM.  While three groups of moir\'e pairs possess sufficient domain curvature (Fig. \ref{fig:TDTG_all_configurations}f-h), only the ABABAB/BCBACA configuration displays calculated spectra that are consistent with our STS results (Fig. \ref{fig:TDTG_all_configurations}k).  The concave domain of this pair (BCBACA) shows a peak at low energy where the convex domain (ABABAB) remains featureless, as in experiment.  In addition, the sharp steps at $\sim\pm350~\textrm{mV}$ in the BCBACA domain, which are associated with the edges of electronic bands, align quantitatively with similar steps observed experimentally in the concave domain (compare Fig. \ref{fig:TDTG_tomography_experiment}g).  This excellent match of the calculated electronic structure with the measured density of states spectrum points conclusively to ABABAB/BCBACA as the stacking configurations spontaneously realized in our TDTG device.  \\

\begin{figure} [htbp]
\centering
\includegraphics[width=6in]{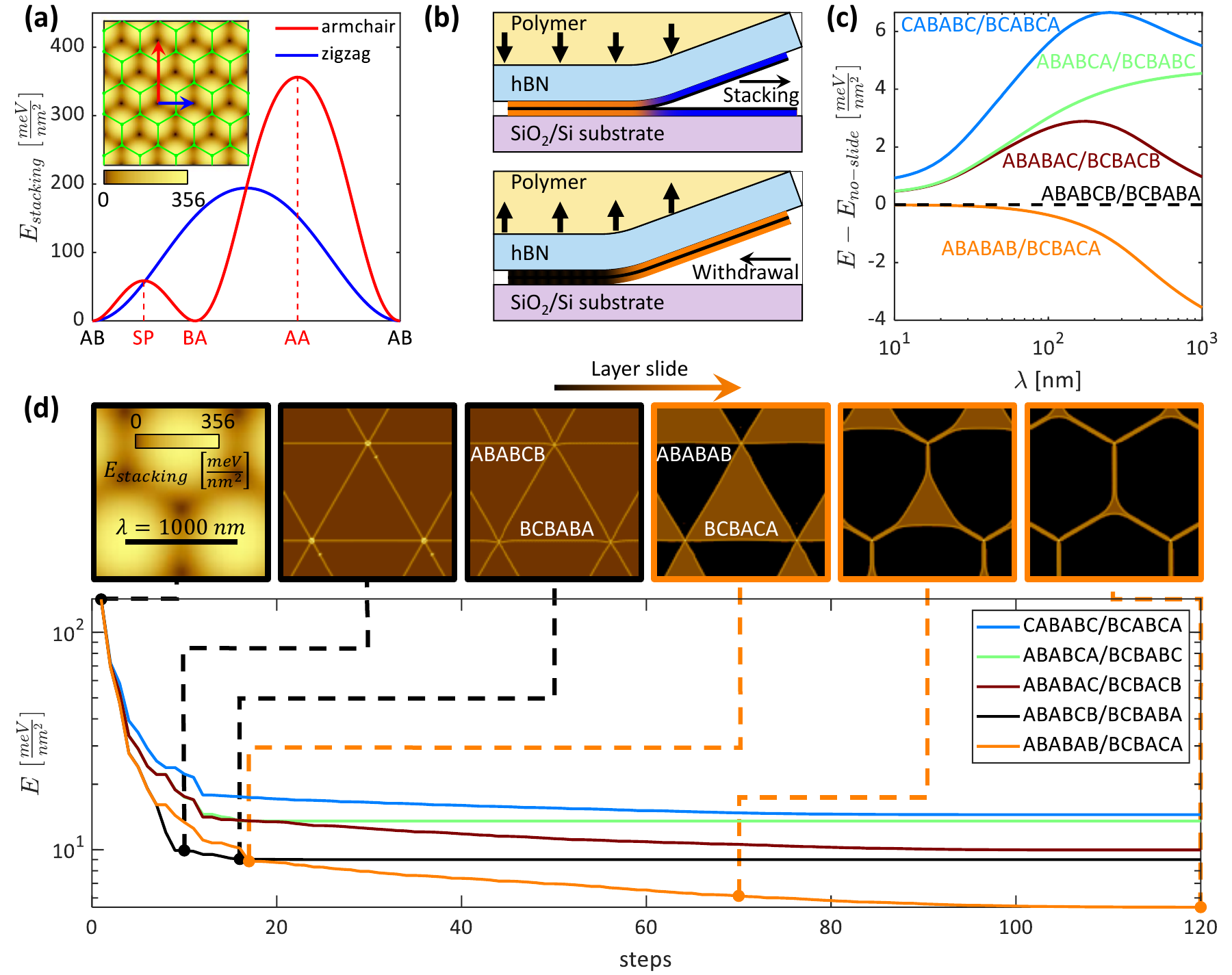}
\caption{\textbf{Energy analysis for atomic relaxation driven formation of unstable phases by sliding layers.} 
\textbf{(a)} Stacking energy density at the interface between two graphene sheets for different configurations. The plot presents cuts along the zigzag and armchair directions. Inset: Stacking energy density in the two-dimensional configuration-space reflecting the stacking energy density as a function of translation of one sheet relative to the other (green honeycomb lattice). \textbf{(b)} Scenarios for generation of a new phase during stacking.  Top: New phase (orange) generation during contact of the two TLGs (blue). Bottom: New phase generation (orange) as the TDTG stack is picked up from the surface.  \textbf{(c)} Comparing total energy density as a function of moir\'e periodicity for different relaxed moir\'e superlattices (see Methods for details on atomic relaxation models). All curves are referenced to the rigid scenario, ABABCB/BCBABA. ABABAB/BCBACA becomes increasingly energetically favorable as the moir\'e period increases. \textbf{(d)} Simulating the ABABCB/BCBABA to ABABAB/BCBACA phase stability inversion through the atomic relaxation process. Each curve shows the total energy density for a particular phase at different optimization steps of the gradient-descent process. Above: instantaneous stacking energy densities of the lowest energy moir\'e system at representative steps. A sharp transition between favorable ABABCB/BCBABA (black) to favorable ABABAB/BCBACA (orange) is observed as the relaxation progresses.}
\label{fig:Sliding_layers_energy}
\end{figure}

Constructing an ABABAB/BCBACA stacking configuration from a minimally twisted Bernal trilayer source crystal requires a global sliding of the middle layer of one of the two twisted trilayers (illustrated in Fig. \ref{fig:TDTG_all_configurations}k, inset).  The energetics of such a global translation are seemingly counter-intuitive because, first, it requires a large energy input to the system to realize a universal layer shift, and second, that shift results in a local increase in the stacking energy density of one of the two domains.  Sliding of a graphene layer can be viewed as continuously traversing the stacking energy landscape shown in Fig. \ref{fig:Sliding_layers_energy}a.  To transform adjacent layers from AB to BA stacking, as required to realize the experimentally observed configuration, the system must cross a formidable energy barrier of  $\sim6\cdot10^4~\frac{\mathrm{eV}}{\mu \mathrm{m}^2}$ set by the saddle-point (SP) of the stacking energy function (indicated by SP in Fig. \ref{fig:Sliding_layers_energy}a).  This cannot conceivably be overcome by thermal excitation alone.  The only step in our experiment during which the sample is subjected to forces of sufficient magnitude to induce a sliding transformation is the stacking process, which involves pressing together each constituent trilayer of the TDTG device before peeling them away from the exfoliation substrate (Fig. \ref{fig:Sliding_layers_energy}b).  When stacking induced sliding transitions like this have been observed in the past\cite{Yang2019}, they have as a rule been from a metastable (rhombohedral) to a stable (Bernal) phase.  In our case, however, the transition from ABABCB/BCBABA (3R/3R) to ABABAB/BCBACA (0R/4R) acts to decrease the thermodynamic stability of nearly half of the device area.\\

Lattice relaxation in TDTG therefore takes an unconventional form in which an energy gain in one half of the crystal is paid for by a relaxation process that occurs in the other.  The energy justification for this phenomenon is studied in Fig. \ref{fig:Sliding_layers_energy}c, where each curve represents the total energy density (stacking and elastic energy) of a given moir\'e pair, after atomic relaxation, as a function of moir\'e wavelength $\lambda$. All curves are referenced to the energy of the rigid configuration (ABABCB/BCBABA), marked by $E_{no-slide}$.  For small $\lambda$ (and therefore weak relaxation) the ABABCB/BCBABA and ABABAB/BCBACA configurations are energetically equivalent, indicating that a layer slide transition without additional relaxation cannot reduce the global energy.  As the moir\'e period increases, the relaxation strengthens, and the energetic imbalance between ABABAB and BCBACA domains (absent in the ABABCB/BCBACA configuration) drives the expansion of the Bernal domain, thus reducing the global energy of the ABABAB/BCBACA configuration relative to its rigid counterpart. If provided sufficient shear forces to overcome the SP energy barrier, the atomic relaxation process can therefore create and stabilize the formation of the otherwise unstable BCBACA phase simply by reducing its relative volume fraction.  \\

We visualize the dynamics of this non-local relaxation in Fig. \ref{fig:Sliding_layers_energy}d, where we plot the instantaneous solution to the energy minimization problem for each stacking considered in Fig. \ref{fig:Sliding_layers_energy}c as a function of iteration number within the steepest-descent optimization algorithm, which can be interpreted as an effective time coordinate, $t$.  At $t=0$, before any relaxation has taken place, all configurations have similar energies.  As the systems flow down their respective energy landscapes, large domains of uniform stacking are formed, separated by straight DWs.  In this intermediate regime, the rigid (ABABCB/BCBABA) and layer slide (ABABAB/BCBACA) scenarios have equivalent energies.  Only when the relaxation process has reached a point where the DWs begin to curve, with the lower energy Bernal phase pushing into the metastable BCBACA configuration, does a layer slide transition become energetically favorable (see crossing of the black and orange lines in Fig. \ref{fig:Sliding_layers_energy}d).  Minimally twisted TDTG thus spontaneously seeks a solution in which a transition to a locally metastable phase (BCBACA) is enabled by a shared phase boundary with a proximate stable structure (ABABAB).\\

The development of new and increasingly complex moir\'e heterostructures demands a revisiting of some of the basic assumptions of van der Waals engineering.  It is sometimes convenient to think of layered materials as immutable building blocks that can be exfoliated and stacked at will.  In reality, however, van der Waals materials inhabit a complicated energy landscape that must be carefully navigated when designing new device architectures.  Moreover, a range of elusive structural phases with exotic electronic properties, such as rhombohedral graphene \cite{Chen2019a,Chen2019b,Zhou2021}, have been difficult to synthesize with traditional techniques.  Our measurements of minimally twisted TDTG reveal a surprising crystallographic transformation that occurs during the stacking process.  The mechanism underlying this transition involves a non-local energy balancing that enables the formation of rhombohedral domains by their coupling to a simultaneously formed relaxed Bernal structure.  Detailed knowledge of this and similar relaxation processes holds the potential to utilize the power of lattice relaxation for engineering novel and otherwise unstable material systems. \\

\section*{Methods}
\subsection*{Samples preparation}
\textbf{Exfoliation:} Graphene and hBN flakes were mechanically exfoliated from the bulk single crystals onto SiO$_2$/Si (285 nm oxide thickness) chips using the tape-assisted exfoliation technique (the tape used was Scotch Magic Tape). The exfoliation followed Ref. \citeonline{Huang2015b}, such that the Si chips were treated with O$_2$ plasma (using a benchtop radio frequency oxygen plasma cleaner of Plasma Etch Inc., PE-50 XL, 100 W at a chamber pressure of ~215 mTorr) for 20 sec for graphene and no O$_2$ plasma treatment for hBN. The chips were then matched with respective exfoliation tape. In the graphene case the chip+tape assembly were heated at $100~^\circ \mathrm{C}$ for 60 sec and cooled to room temperature prior to removing the tape. Such thermal treatment was not done for hBN. \\

\noindent
\textbf{Stack preparation:} The heterostructure was assembled using standard dry-transfer techniques\cite{Wang2013b} with a polypropylene carbonate (PPC) film mounted on a transparent-tape-covered polydimethylsiloxane (PDMS) stamp. The transparent tape layer was added to the stamp to mold the PDMS into a hemispherical shape which provides precise control of the PPC contact area during assembly\cite{Kim2016b}. The heterostructure was made by first picking up the hBN (20 nm thick). Prior to pick-up, mechanically exfoliated TLG flakes on Si/SiO$_2$ were separately patterned with anodic-oxidation lithography\cite{Li2018} to facilitate the "cut-and-stack" technique\cite{Saito2020b}. 
Next, the PPC film with the heterostructure on top is mechanically removed from the transparent-tape-covered PDMS stamp and placed onto a Si/SiO$_2$ substrate such that the final pick-up layer is the top layer. Then the underlying PPC was removed by vacuum annealing at $350~^\circ$C. 
\subsection*{Near-field imaging techniques}
The mid-IR near-field scans in this work were acquired with a phase-resolved scattering type scanning optical microscope imaging (s-SNOM) with a commercial system (Neaspec), using a mid-IR quantum cascade laser (Hedgehog by Daylight Solutions) tuned between $8.7-10.2~\mu \mathrm{m}$. The laser light was focused to a diffraction limited spot at the apex of a metallic tip, while raster scanning the sample at tapping mode. We collect the scattered light (power of $3-5~\mathrm{mW}$) by a cryogenic HgCdTe detector (Kolmar Technologies). The near-field amplitude and phase were extracted as harmonic components of the tapping frequency using an interferometric detection method, the pseudo-heterodyne scheme, by interfering the scattered light with a modulated reference arm at the detector\cite{Sunku2018h}. The near-field scans of Fig. \ref{fig:TDTG_overview} and Fig. \ref{fig:TDTG_tomography_experiment} were taken at 983 and 1000 $\mathrm{cm}^{-1}$ respectively. 

\subsection*{Scanning tunneling microscopy and spectroscopy}
STM/S measurements were conducted in a home-built STM under ultra-high vacuum at 7K.  The tungsten tunneling tip was electrochemically etched and calibrated against the Au(111) surface state prior to each sample approach.  Spectroscopy was measured using a lock-in amplifier to record the differential conductance with a bias modulation between 1-7 mV at 927 Hz, a set point voltage of 250 mV, and a set point current of 120 pA.

\subsection*{Electronic structure theory calculations of generalized stacking fault energy function (GSFE) and DOS}

In order to capture the generalized stacking fault energy function (GSFE) and electron density of states (DOS), we rely on density functional theory calculations. The different six-layer graphene stacking configurations were captured within a hexagonal unit cell with an in-plane lattice constant of $a = b = 2.459$~\AA. We describe the system using a $24\times24\times1$ k-point mesh within the plane-wave code JDFTx~\cite{sundararaman_jdftx_2017}. Fermi smearing of width 0.01 Hartree is applied to the electronic occupations. 
We use ultrasoft pseudopotentials~\cite{garrity_pseudopotentials_2014} and the PBEsol exchange-correlation functional~\cite{perdew_restoring_2008}. In order to remove any artificial interactions between periodic images in the out-of-plane direction, we use a Coulomb truncation technique~\cite{sundararaman_regularization_2013}. The plane-wave cutoff used is 40 Hartrees. We use a stringent charge density cutoff of 1000 Hartrees in order to densely sample the $z$ direction of the unit cell, which is important for accurately describing the energetics of the different stacking configurations and therefore for comparison among them. Van der Waals interactions between the graphene layers are modeled using the DFT-D3 scheme~\cite{grimme_consistent_2010}.\\

Using these calculations as starting points, we can then capture the electronic density of states. We describe the electronic states of our systems using a real-space Wannier representation based on maximally-localized Wannier functions~\cite{marzari_maximally_1997}. This allows us to sample the electronic energies at arbitrary wave vectors. In our DOS calculations, we sample $5.76\times10^{7}$ wave vectors to accurately converge the DOS. We use a Lorentzian with a broadening of width 4.3~meV to smooth the results.

\subsection*{Atomic relaxation calculations}
Modeling of the atomic relaxation of TDTG structures was performed within a continuity model
following the method presented in Refs. \citeonline{carr:2018:relaxation,Halbertal2021}. In this model, the total energy of the system is taken as the sum of elastic energy and a stacking energy term. The total energy was minimized in search for the inter-layer real space displacement field corresponding to the relaxed structure. The stacking configuration at the interface was imposed to be AA at the four corners of the moir\'e unit-cell. The mechanical relaxation parameters (bulk and shear moduli) for TLG as well as the generalized stacking fault energy function (GSFE) for the TLG/TLG interface were calculated using DFT (see DFT section in Methods for details).
The resulting mechanical coefficients for TLG (in meV per unit-cell) are: bulk modulus - $K$=210971, shear modulus - $G$=151580.\\

The GSFE coefficients were extracted from a $7 \times 7$ sampling of the configuration between two $ABA$-TLG, with the vertical positions of the atoms relaxed at each configuration.
The Fourier components of the resulting energies were then extracted to create a convenient functional form for the GSFE used to describe the stacking energy term in the atomic relaxation calculations. For simplicity the GSFE for configurations other than the nominal case (ABABCB/BCBABA) used the extracted GSFE for the nominal case while imposing the stacking energies at the lowest energy configurations as calculated by DFT (values are detailed in Table \ref{tab:E_lowest_conf}). The GSFE for a given configuration was taken as the closest curve ($L_2$ norm) with the same functional structure, that satisfies the imposed lowest energy configuration. This approach circumvented the need to calculate the full stacking landscape for all systems. The validity of this approach was assessed comparing the resulting GSFE with the full GSFE calculation for BCABCA/CABABC yielding similar results. The resulting GSFE coefficients are detailed in Table \ref{tab:material_param}.

\bibliography{TDTG_bibliography}

\begin{thebibliography}{10}
\urlstyle{rm}
\expandafter\ifx\csname url\endcsname\relax
  \def\url#1{\texttt{#1}}\fi
\expandafter\ifx\csname urlprefix\endcsname\relax\def\urlprefix{URL }\fi
\expandafter\ifx\csname doiprefix\endcsname\relax\def\doiprefix{DOI: }\fi
\providecommand{\bibinfo}[2]{#2}
\providecommand{\eprint}[2][]{\url{#2}}

\bibitem{Cao2018_insulator}
\bibinfo{author}{Cao, Y.} \emph{et~al.}
\newblock \bibinfo{journal}{\bibinfo{title}{Correlated insulator behaviour at
  half-filling in magic-angle graphene superlattices}}.
\newblock {\emph{\JournalTitle{Nature}}} \textbf{\bibinfo{volume}{556}},
  \bibinfo{pages}{80--84}, \doiprefix\url{10.1038/nature26154}
  (\bibinfo{year}{2018}).

\bibitem{Cao2018_superconductor}
\bibinfo{author}{Cao, Y.} \emph{et~al.}
\newblock \bibinfo{journal}{\bibinfo{title}{Unconventional superconductivity in
  magic-angle graphene superlattices}}.
\newblock {\emph{\JournalTitle{Nature}}} \textbf{\bibinfo{volume}{556}},
  \bibinfo{pages}{43--50}, \doiprefix\url{10.1038/nature26160}
  (\bibinfo{year}{2018}).

\bibitem{Xie2021}
\bibinfo{author}{Xie, Y.} \emph{et~al.}
\newblock \bibinfo{journal}{\bibinfo{title}{Fractional chern insulators in
  magic-angle twisted bilayer graphene}}.
\newblock {\emph{\JournalTitle{Nature}}} \textbf{\bibinfo{volume}{600}},
  \bibinfo{pages}{439--443}, \doiprefix\url{10.1038/s41586-021-04002-3}
  (\bibinfo{year}{2021}).

\bibitem{Nuckolls2020}
\bibinfo{author}{Nuckolls, K.~P.} \emph{et~al.}
\newblock \bibinfo{journal}{\bibinfo{title}{Strongly correlated chern
  insulators in magic-angle twisted bilayer graphene}}.
\newblock {\emph{\JournalTitle{Nature}}} \textbf{\bibinfo{volume}{588}},
  \bibinfo{pages}{610--615}, \doiprefix\url{10.1038/s41586-020-3028-8}
  (\bibinfo{year}{2020}).

\bibitem{Liu2020}
\bibinfo{author}{Liu, X.} \emph{et~al.}
\newblock \bibinfo{journal}{\bibinfo{title}{Tunable spin-polarized correlated
  states in twisted double bilayer graphene}}.
\newblock {\emph{\JournalTitle{Nature}}} \textbf{\bibinfo{volume}{583}},
  \bibinfo{pages}{221--225}, \doiprefix\url{10.1038/s41586-020-2458-7}
  (\bibinfo{year}{2020}).

\bibitem{Cao2020_tdbg}
\bibinfo{author}{Cao, Y.} \emph{et~al.}
\newblock \bibinfo{journal}{\bibinfo{title}{Tunable correlated states and
  spin-polarized phases in twisted bilayer--bilayer graphene}}.
\newblock {\emph{\JournalTitle{Nature}}} \textbf{\bibinfo{volume}{583}},
  \bibinfo{pages}{215--220}, \doiprefix\url{10.1038/s41586-020-2260-6}
  (\bibinfo{year}{2020}).

\bibitem{He2021}
\bibinfo{author}{He, M.} \emph{et~al.}
\newblock \bibinfo{journal}{\bibinfo{title}{Symmetry breaking in twisted double
  bilayer graphene}}.
\newblock {\emph{\JournalTitle{Nature Physics}}} \textbf{\bibinfo{volume}{17}},
  \bibinfo{pages}{26--30}, \doiprefix\url{10.1038/s41567-020-1030-6}
  (\bibinfo{year}{2021}).

\bibitem{PhysRevLett.123.197702}
\bibinfo{author}{Burg, G.~W.} \emph{et~al.}
\newblock \bibinfo{journal}{\bibinfo{title}{Correlated insulating states in
  twisted double bilayer graphene}}.
\newblock {\emph{\JournalTitle{Phys. Rev. Lett.}}}
  \textbf{\bibinfo{volume}{123}}, \bibinfo{pages}{197702},
  \doiprefix\url{10.1103/PhysRevLett.123.197702} (\bibinfo{year}{2019}).

\bibitem{Ghiotto2021}
\bibinfo{author}{Ghiotto, A.} \emph{et~al.}
\newblock \bibinfo{journal}{\bibinfo{title}{Quantum criticality in twisted
  transition metal dichalcogenides}}.
\newblock {\emph{\JournalTitle{Nature}}} \textbf{\bibinfo{volume}{597}},
  \bibinfo{pages}{345--349}, \doiprefix\url{10.1038/s41586-021-03815-6}
  (\bibinfo{year}{2021}).

\bibitem{Jaoui2022}
\bibinfo{author}{Jaoui, A.} \emph{et~al.}
\newblock \bibinfo{journal}{\bibinfo{title}{Quantum critical behaviour in
  magic-angle twisted bilayer graphene}}.
\newblock {\emph{\JournalTitle{Nature Physics}}}
  \doiprefix\url{10.1038/s41567-022-01556-5} (\bibinfo{year}{2022}).

\bibitem{Rubio-Verd2022}
\bibinfo{author}{Rubio-Verd{\'u}, C.} \emph{et~al.}
\newblock \bibinfo{journal}{\bibinfo{title}{Moir{\'e} nematic phase in twisted
  double bilayer graphene}}.
\newblock {\emph{\JournalTitle{Nature Physics}}} \textbf{\bibinfo{volume}{18}},
  \bibinfo{pages}{196--202}, \doiprefix\url{10.1038/s41567-021-01438-2}
  (\bibinfo{year}{2022}).

\bibitem{Samajdar_2021}
\bibinfo{author}{Samajdar, R.} \emph{et~al.}
\newblock \bibinfo{journal}{\bibinfo{title}{Electric-field-tunable electronic
  nematic order in twisted double-bilayer graphene}}.
\newblock {\emph{\JournalTitle{2D Materials}}} \textbf{\bibinfo{volume}{8}},
  \bibinfo{pages}{034005}, \doiprefix\url{10.1088/2053-1583/abfcd6}
  (\bibinfo{year}{2021}).

\bibitem{Kerelsky2019}
\bibinfo{author}{Kerelsky, A.} \emph{et~al.}
\newblock \bibinfo{journal}{\bibinfo{title}{Maximized electron interactions at
  the magic angle in twisted bilayer graphene}}.
\newblock {\emph{\JournalTitle{Nature}}} \textbf{\bibinfo{volume}{572}},
  \bibinfo{pages}{95--100}, \doiprefix\url{10.1038/s41586-019-1431-9}
  (\bibinfo{year}{2019}).

\bibitem{Jiang2019}
\bibinfo{author}{Jiang, Y.} \emph{et~al.}
\newblock \bibinfo{journal}{\bibinfo{title}{Charge order and broken rotational
  symmetry in magic-angle twisted bilayer graphene}}.
\newblock {\emph{\JournalTitle{Nature}}} \textbf{\bibinfo{volume}{573}},
  \bibinfo{pages}{91--95}, \doiprefix\url{10.1038/s41586-019-1460-4}
  (\bibinfo{year}{2019}).

\bibitem{Li2021}
\bibinfo{author}{Li, H.} \emph{et~al.}
\newblock \bibinfo{journal}{\bibinfo{title}{Imaging two-dimensional generalized
  wigner crystals}}.
\newblock {\emph{\JournalTitle{Nature}}} \textbf{\bibinfo{volume}{597}},
  \bibinfo{pages}{650--654}, \doiprefix\url{10.1038/s41586-021-03874-9}
  (\bibinfo{year}{2021}).

\bibitem{Kennes2021}
\bibinfo{author}{Kennes, D.~M.} \emph{et~al.}
\newblock \bibinfo{journal}{\bibinfo{title}{Moir{\'e} heterostructures as a
  condensed-matter quantum simulator}}.
\newblock {\emph{\JournalTitle{Nature Physics}}} \textbf{\bibinfo{volume}{17}},
  \bibinfo{pages}{155--163}, \doiprefix\url{10.1038/s41567-020-01154-3}
  (\bibinfo{year}{2021}).

\bibitem{doi:10.1126/science.abg0399}
\bibinfo{author}{Hao, Z.} \emph{et~al.}
\newblock \bibinfo{journal}{\bibinfo{title}{Electric field\&\#x2013;tunable
  superconductivity in alternating-twist magic-angle trilayer graphene}}.
\newblock {\emph{\JournalTitle{Science}}} \textbf{\bibinfo{volume}{371}},
  \bibinfo{pages}{1133--1138}, \doiprefix\url{10.1126/science.abg0399}
  (\bibinfo{year}{2021}).
\newblock \eprint{https://www.science.org/doi/pdf/10.1126/science.abg0399}.

\bibitem{Park2021}
\bibinfo{author}{Park, J.~M.}, \bibinfo{author}{Cao, Y.},
  \bibinfo{author}{Watanabe, K.}, \bibinfo{author}{Taniguchi, T.} \&
  \bibinfo{author}{Jarillo-Herrero, P.}
\newblock \bibinfo{journal}{\bibinfo{title}{Tunable strongly coupled
  superconductivity in magic-angle twisted trilayer graphene}}.
\newblock {\emph{\JournalTitle{Nature}}} \textbf{\bibinfo{volume}{590}},
  \bibinfo{pages}{249--255}, \doiprefix\url{10.1038/s41586-021-03192-0}
  (\bibinfo{year}{2021}).

\bibitem{2112.07127}
\bibinfo{author}{Siriviboon, P.} \emph{et~al.}
\newblock \bibinfo{title}{A new flavor of correlation and superconductivity in
  small twist-angle trilayer graphene} (\bibinfo{year}{2021}).
\newblock \eprint{arXiv:2112.07127}.

\bibitem{2112.07841}
\bibinfo{author}{Lin, J.-X.} \emph{et~al.}
\newblock \bibinfo{title}{Zero-field superconducting diode effect in
  small-twist-angle trilayer graphene} (\bibinfo{year}{2021}).
\newblock \eprint{arXiv:2112.07841}.

\bibitem{Turkel2022}
\bibinfo{author}{Turkel, S.} \emph{et~al.}
\newblock \bibinfo{journal}{\bibinfo{title}{Orderly disorder in magic-angle
  twisted trilayer graphene}}.
\newblock {\emph{\JournalTitle{Science}}} \textbf{\bibinfo{volume}{376}},
  \bibinfo{pages}{193--199}, \doiprefix\url{10.1126/science.abk1895}
  (\bibinfo{year}{2022}).
\newblock \eprint{https://www.science.org/doi/pdf/10.1126/science.abk1895}.

\bibitem{Halbertal2021}
\bibinfo{author}{Halbertal, D.} \emph{et~al.}
\newblock \bibinfo{journal}{\bibinfo{title}{{Moir{\'{e}} metrology of energy
  landscapes in van der Waals heterostructures}}}.
\newblock {\emph{\JournalTitle{Nature Communications}}}
  \textbf{\bibinfo{volume}{12}}, \bibinfo{pages}{1--8},
  \doiprefix\url{10.1038/s41467-020-20428-1} (\bibinfo{year}{2021}).

\bibitem{Kerelsky2021}
\bibinfo{author}{Kerelsky, A.} \emph{et~al.}
\newblock \bibinfo{journal}{\bibinfo{title}{{Erratum: Moir{\'{e}}less
  correlations in ABCA graphene (Proceedings of the National Academy of
  Sciences of the United States of America (2021) 118 (e2017366118) DOI:
  10.1073/pnas.2017366118)}}}.
\newblock {\emph{\JournalTitle{Proceedings of the National Academy of Sciences
  of the United States of America}}} \textbf{\bibinfo{volume}{118}},
  \doiprefix\url{10.1073/pnas.2102335118} (\bibinfo{year}{2021}).

\bibitem{PhysRevB.96.075311}
\bibinfo{author}{Nam, N. N.~T.} \& \bibinfo{author}{Koshino, M.}
\newblock \bibinfo{journal}{\bibinfo{title}{Lattice relaxation and energy band
  modulation in twisted bilayer graphene}}.
\newblock {\emph{\JournalTitle{Phys. Rev. B}}} \textbf{\bibinfo{volume}{96}},
  \bibinfo{pages}{075311}, \doiprefix\url{10.1103/PhysRevB.96.075311}
  (\bibinfo{year}{2017}).

\bibitem{Haddadi2020}
\bibinfo{author}{Haddadi, F.}, \bibinfo{author}{Wu, Q.},
  \bibinfo{author}{Kruchkov, A.~J.} \& \bibinfo{author}{Yazyev, O.~V.}
\newblock \bibinfo{journal}{\bibinfo{title}{Moir{\'e} flat bands in twisted
  double bilayer graphene}}.
\newblock {\emph{\JournalTitle{Nano Letters}}} \textbf{\bibinfo{volume}{20}},
  \bibinfo{pages}{2410--2415}, \doiprefix\url{10.1021/acs.nanolett.9b05117}
  (\bibinfo{year}{2020}).

\bibitem{PhysRevB.99.205134}
\bibinfo{author}{Guinea, F.} \& \bibinfo{author}{Walet, N.~R.}
\newblock \bibinfo{journal}{\bibinfo{title}{Continuum models for twisted
  bilayer graphene: Effect of lattice deformation and hopping parameters}}.
\newblock {\emph{\JournalTitle{Phys. Rev. B}}} \textbf{\bibinfo{volume}{99}},
  \bibinfo{pages}{205134}, \doiprefix\url{10.1103/PhysRevB.99.205134}
  (\bibinfo{year}{2019}).

\bibitem{moore_nanoscale_2021}
\bibinfo{author}{Moore, S.~L.} \emph{et~al.}
\newblock \bibinfo{journal}{\bibinfo{title}{Nanoscale lattice dynamics in
  hexagonal boron nitride moiré superlattices}}.
\newblock {\emph{\JournalTitle{Nature Communications}}}
  \textbf{\bibinfo{volume}{12}}, \bibinfo{pages}{5741},
  \doiprefix\url{10.1038/s41467-021-26072-7} (\bibinfo{year}{2021}).

\bibitem{Sunku2018h}
\bibinfo{author}{Sunku, S.~S.} \emph{et~al.}
\newblock \bibinfo{journal}{\bibinfo{title}{{Photonic crystals for nano-light
  in moir{\'{e}} graphene superlattices}}}.
\newblock {\emph{\JournalTitle{Science}}} \textbf{\bibinfo{volume}{362}},
  \bibinfo{pages}{1153--1156}, \doiprefix\url{10.1126/science.aau5144}
  (\bibinfo{year}{2018}).

\bibitem{Hesp2021}
\bibinfo{author}{Hesp, N.~C.} \emph{et~al.}
\newblock \bibinfo{journal}{\bibinfo{title}{{Observation of interband
  collective excitations in twisted bilayer graphene}}}.
\newblock {\emph{\JournalTitle{Nature Physics}}} \textbf{\bibinfo{volume}{17}},
  \bibinfo{pages}{1162--1168}, \doiprefix\url{10.1038/s41567-021-01327-8}
  (\bibinfo{year}{2021}).

\bibitem{Jiang2016a}
\bibinfo{author}{Jiang, L.} \emph{et~al.}
\newblock \bibinfo{journal}{\bibinfo{title}{{Soliton-dependent plasmon
  reflection at bilayer graphene domain walls}}}.
\newblock {\emph{\JournalTitle{Nature Materials}}}
  \textbf{\bibinfo{volume}{15}}, \bibinfo{pages}{840--844},
  \doiprefix\url{10.1038/nmat4653} (\bibinfo{year}{2016}).

\bibitem{Yang2019}
\bibinfo{author}{Yang, Y.} \& \bibinfo{author}{Al, E.}
\newblock \bibinfo{journal}{\bibinfo{title}{{Stacking Order in Graphite Films
  Controlled by van der Waals Technology}}}.
\newblock {\emph{\JournalTitle{Nano Letters}}}
  \doiprefix\url{10.1021/acs.nanolett.9b03014} (\bibinfo{year}{2019}).

\bibitem{Chen2019a}
\bibinfo{author}{Chen, G.} \emph{et~al.}
\newblock \bibinfo{journal}{\bibinfo{title}{Signatures of tunable
  superconductivity in a trilayer graphene moir{\'e} superlattice}}.
\newblock {\emph{\JournalTitle{Nature}}} \textbf{\bibinfo{volume}{572}},
  \bibinfo{pages}{215--219}, \doiprefix\url{10.1038/s41586-019-1393-y}
  (\bibinfo{year}{2019}).

\bibitem{Chen2019b}
\bibinfo{author}{Chen, G.} \emph{et~al.}
\newblock \bibinfo{journal}{\bibinfo{title}{Evidence of a gate-tunable mott
  insulator in a trilayer graphene moir{\'e} superlattice}}.
\newblock {\emph{\JournalTitle{Nature Physics}}} \textbf{\bibinfo{volume}{15}},
  \bibinfo{pages}{237--241}, \doiprefix\url{10.1038/s41567-018-0387-2}
  (\bibinfo{year}{2019}).

\bibitem{Zhou2021}
\bibinfo{author}{Zhou, H.}, \bibinfo{author}{Xie, T.},
  \bibinfo{author}{Taniguchi, T.}, \bibinfo{author}{Watanabe, K.} \&
  \bibinfo{author}{Young, A.~F.}
\newblock \bibinfo{journal}{\bibinfo{title}{Superconductivity in rhombohedral
  trilayer graphene}}.
\newblock {\emph{\JournalTitle{Nature}}} \textbf{\bibinfo{volume}{598}},
  \bibinfo{pages}{434--438}, \doiprefix\url{10.1038/s41586-021-03926-0}
  (\bibinfo{year}{2021}).

\bibitem{Huang2015b}
\bibinfo{author}{Huang, Y.} \emph{et~al.}
\newblock \bibinfo{journal}{\bibinfo{title}{{Reliable Exfoliation of Large-Area
  High-Quality Flakes of Graphene and Other Two-Dimensional Materials}}}.
\newblock {\emph{\JournalTitle{ACS Nano}}} \textbf{\bibinfo{volume}{9}},
  \bibinfo{pages}{10612--10620}, \doiprefix\url{10.1021/acsnano.5b04258}
  (\bibinfo{year}{2015}).

\bibitem{Wang2013b}
\bibinfo{author}{Wang, L.} \emph{et~al.}
\newblock \bibinfo{journal}{\bibinfo{title}{{One-Dimensional Electrical Contact
  to a Two-Dimensional Material}}}.
\newblock {\emph{\JournalTitle{Science}}} \textbf{\bibinfo{volume}{342}},
  \bibinfo{pages}{614--618}, \doiprefix\url{10.1126/science.1244358}
  (\bibinfo{year}{2013}).

\bibitem{Kim2016b}
\bibinfo{author}{Kim, K.} \emph{et~al.}
\newblock \bibinfo{journal}{\bibinfo{title}{{Van der Waals Heterostructures
  with High Accuracy Rotational Alignment}}}.
\newblock {\emph{\JournalTitle{Nano Letters}}} \textbf{\bibinfo{volume}{16}},
  \bibinfo{pages}{1989--1995}, \doiprefix\url{10.1021/acs.nanolett.5b05263}
  (\bibinfo{year}{2016}).

\bibitem{Li2018}
\bibinfo{author}{Li, H.} \emph{et~al.}
\newblock \bibinfo{journal}{\bibinfo{title}{{Electrode-Free Anodic Oxidation
  Nanolithography of Low-Dimensional Materials}}}.
\newblock {\emph{\JournalTitle{Nano Letters}}} \textbf{\bibinfo{volume}{18}},
  \bibinfo{pages}{8011--8015}, \doiprefix\url{10.1021/acs.nanolett.8b04166}
  (\bibinfo{year}{2018}).

\bibitem{Saito2020b}
\bibinfo{author}{Saito, Y.}, \bibinfo{author}{Ge, J.},
  \bibinfo{author}{Watanabe, K.}, \bibinfo{author}{Taniguchi, T.} \&
  \bibinfo{author}{Young, A.~F.}
\newblock \bibinfo{journal}{\bibinfo{title}{{Independent superconductors and
  correlated insulators in twisted bilayer graphene}}}.
\newblock {\emph{\JournalTitle{Nature Physics}}} \textbf{\bibinfo{volume}{16}},
  \bibinfo{pages}{926--930}, \doiprefix\url{10.1038/s41567-020-0928-3}
  (\bibinfo{year}{2020}).

\bibitem{sundararaman_jdftx_2017}
\bibinfo{author}{Sundararaman, R.} \emph{et~al.}
\newblock \bibinfo{journal}{\bibinfo{title}{{JDFTx}: {Software} for joint
  density-functional theory}}.
\newblock {\emph{\JournalTitle{SoftwareX}}} \textbf{\bibinfo{volume}{6}},
  \bibinfo{pages}{278--284}, \doiprefix\url{10.1016/j.softx.2017.10.006}
  (\bibinfo{year}{2017}).

\bibitem{garrity_pseudopotentials_2014}
\bibinfo{author}{Garrity, K.~F.}, \bibinfo{author}{Bennett, J.~W.},
  \bibinfo{author}{Rabe, K.~M.} \& \bibinfo{author}{Vanderbilt, D.}
\newblock \bibinfo{journal}{\bibinfo{title}{Pseudopotentials for
  high-throughput {DFT} calculations}}.
\newblock {\emph{\JournalTitle{Computational Materials Science}}}
  \textbf{\bibinfo{volume}{81}}, \bibinfo{pages}{446--452},
  \doiprefix\url{10.1016/j.commatsci.2013.08.053} (\bibinfo{year}{2014}).

\bibitem{perdew_restoring_2008}
\bibinfo{author}{Perdew, J.~P.} \emph{et~al.}
\newblock \bibinfo{journal}{\bibinfo{title}{Restoring the {Density}-{Gradient}
  {Expansion} for {Exchange} in {Solids} and {Surfaces}}}.
\newblock {\emph{\JournalTitle{Physical Review Letters}}}
  \textbf{\bibinfo{volume}{100}}, \bibinfo{pages}{136406},
  \doiprefix\url{10.1103/PhysRevLett.100.136406} (\bibinfo{year}{2008}).

\bibitem{sundararaman_regularization_2013}
\bibinfo{author}{Sundararaman, R.} \& \bibinfo{author}{Arias, T.~A.}
\newblock \bibinfo{journal}{\bibinfo{title}{Regularization of the {Coulomb}
  singularity in exact exchange by {Wigner}-{Seitz} truncated interactions:
  {Towards} chemical accuracy in nontrivial systems}}.
\newblock {\emph{\JournalTitle{Physical Review B}}}
  \textbf{\bibinfo{volume}{87}}, \bibinfo{pages}{165122},
  \doiprefix\url{10.1103/PhysRevB.87.165122} (\bibinfo{year}{2013}).

\bibitem{grimme_consistent_2010}
\bibinfo{author}{Grimme, S.}, \bibinfo{author}{Antony, J.},
  \bibinfo{author}{Ehrlich, S.} \& \bibinfo{author}{Krieg, H.}
\newblock \bibinfo{journal}{\bibinfo{title}{A consistent and accurate ab initio
  parametrization of density functional dispersion correction ({DFT}-{D}) for
  the 94 elements {H}-{Pu}}}.
\newblock {\emph{\JournalTitle{The Journal of Chemical Physics}}}
  \textbf{\bibinfo{volume}{132}}, \bibinfo{pages}{154104},
  \doiprefix\url{10.1063/1.3382344} (\bibinfo{year}{2010}).

\bibitem{marzari_maximally_1997}
\bibinfo{author}{Marzari, N.} \& \bibinfo{author}{Vanderbilt, D.}
\newblock \bibinfo{journal}{\bibinfo{title}{Maximally localized generalized
  {Wannier} functions for composite energy bands}}.
\newblock {\emph{\JournalTitle{Physical Review B}}}
  \textbf{\bibinfo{volume}{56}}, \bibinfo{pages}{12847--12865},
  \doiprefix\url{10.1103/PhysRevB.56.12847} (\bibinfo{year}{1997}).

\bibitem{carr:2018:relaxation}
\bibinfo{author}{Carr, S.} \emph{et~al.}
\newblock \bibinfo{journal}{\bibinfo{title}{Relaxation and domain formation in
  incommensurate two-dimensional heterostructures}}.
\newblock {\emph{\JournalTitle{Physical Review B}}}
  \textbf{\bibinfo{volume}{98}}, \bibinfo{pages}{224102}
  (\bibinfo{year}{2018}).

\bibitem{lui_imaging_2011}
\bibinfo{author}{Lui, C.~H.} \emph{et~al.}
\newblock \bibinfo{journal}{\bibinfo{title}{Imaging {Stacking} {Order} in
  {Few}-{Layer} {Graphene}}}.
\newblock {\emph{\JournalTitle{Nano Letters}}} \textbf{\bibinfo{volume}{11}},
  \bibinfo{pages}{164--169}, \doiprefix\url{10.1021/nl1032827}
  (\bibinfo{year}{2011}).

\bibitem{mak_evolution_2010}
\bibinfo{author}{Mak, K.~F.}, \bibinfo{author}{Sfeir, M.~Y.},
  \bibinfo{author}{Misewich, J.~A.} \& \bibinfo{author}{Heinz, T.~F.}
\newblock \bibinfo{journal}{\bibinfo{title}{The evolution of electronic
  structure in few-layer graphene revealed by optical spectroscopy}}.
\newblock {\emph{\JournalTitle{Proceedings of the National Academy of
  Sciences}}} \textbf{\bibinfo{volume}{107}}, \bibinfo{pages}{14999--15004},
  \doiprefix\url{10.1073/pnas.1004595107} (\bibinfo{year}{2010}).

\bibitem{kubo_statistical-mechanical_1957}
\bibinfo{author}{Kubo, R.}
\newblock \bibinfo{journal}{\bibinfo{title}{Statistical-{Mechanical} {Theory}
  of {Irreversible} {Processes}. {I}. {General} {Theory} and {Simple}
  {Applications} to {Magnetic} and {Conduction} {Problems}}}.
\newblock {\emph{\JournalTitle{Journal of the Physical Society of Japan}}}
  \textbf{\bibinfo{volume}{12}}, \bibinfo{pages}{570--586},
  \doiprefix\url{10.1143/JPSJ.12.570} (\bibinfo{year}{1957}).
\newblock \bibinfo{note}{\_eprint: https://doi.org/10.1143/JPSJ.12.570}.

\bibitem{Cong2011}
\bibinfo{author}{Cong, C.} \emph{et~al.}
\newblock \bibinfo{journal}{\bibinfo{title}{Raman characterization of aba- and
  abc-stacked trilayer graphene}}.
\newblock {\emph{\JournalTitle{ACS Nano}}} \textbf{\bibinfo{volume}{5}},
  \bibinfo{pages}{8760--8768}, \doiprefix\url{10.1021/nn203472f}
  (\bibinfo{year}{2011}).

\bibitem{Sun2009}
\bibinfo{author}{Sun, J.}, \bibinfo{author}{Schotland, J.~C.},
  \bibinfo{author}{Hillenbrand, R.} \& \bibinfo{author}{Carney, P.~S.}
\newblock \bibinfo{journal}{\bibinfo{title}{{Nanoscale optical tomography using
  volume-scanning near-field microscopy}}}.
\newblock {\emph{\JournalTitle{Applied Physics Letters}}}
  \textbf{\bibinfo{volume}{95}}, \doiprefix\url{10.1063/1.3224177}
  (\bibinfo{year}{2009}).

\bibitem{Govyadinov2014b}
\bibinfo{author}{Govyadinov, A.~A.} \emph{et~al.}
\newblock \bibinfo{journal}{\bibinfo{title}{{Recovery of permittivity and depth
  from near-field data as a step toward infrared nanotomography}}}.
\newblock {\emph{\JournalTitle{ACS Nano}}} \textbf{\bibinfo{volume}{8}},
  \bibinfo{pages}{6911--6921}, \doiprefix\url{10.1021/nn5016314}
  (\bibinfo{year}{2014}).

\bibitem{Mooshammer2018}
\bibinfo{author}{Mooshammer, F.} \emph{et~al.}
\newblock \bibinfo{journal}{\bibinfo{title}{{Nanoscale Near-Field Tomography of
  Surface States on (Bi 0.5 Sb 0.5 ) 2 Te 3}}}.
\newblock {\emph{\JournalTitle{Nano Letters}}} \textbf{\bibinfo{volume}{18}},
  \bibinfo{pages}{7515--7523}, \doiprefix\url{10.1021/acs.nanolett.8b03008}
  (\bibinfo{year}{2018}).

\bibitem{Taubner2005}
\bibinfo{author}{Taubner, T.}, \bibinfo{author}{Keilmann, F.} \&
  \bibinfo{author}{Hillenbrand, R.}
\newblock \bibinfo{journal}{\bibinfo{title}{{Nanoscale-resolved subsurface
  imaging by scattering-type near-field optical microscopy}}}.
\newblock {\emph{\JournalTitle{Optics Express}}} \textbf{\bibinfo{volume}{13}},
  \bibinfo{pages}{8893}, \doiprefix\url{10.1364/opex.13.008893}
  (\bibinfo{year}{2005}).

\bibitem{PhysRevB.90.085136}
\bibinfo{author}{McLeod, A.~S.} \emph{et~al.}
\newblock \bibinfo{journal}{\bibinfo{title}{Model for quantitative tip-enhanced
  spectroscopy and the extraction of nanoscale-resolved optical constants}}.
\newblock {\emph{\JournalTitle{Phys. Rev. B}}} \textbf{\bibinfo{volume}{90}},
  \bibinfo{pages}{085136}, \doiprefix\url{10.1103/PhysRevB.90.085136}
  (\bibinfo{year}{2014}).

\end{thebibliography}

\section*{Acknowledgements}
Nano-imaging research at Columbia is supported by DOE-BES grant DE-SC0018426. STM measurements were supported by the Office of Basic Energy Sciences, Materials Sciences and Engineering Division, U.S. Department of Energy (DOE) under Contract No. DE-SC0012704. ANP acknowledges salary support from the National Science Foundation via grant DMR-2004691. Research on atomic relaxation is supported by W911NF2120147. Work by C.J.C. and P.N. was primarily supported by the Department of Energy, Photonics at Thermodynamic Limits Energy Frontier Research Center, under Grant No. DE-SC0019140. We  acknowledge  funding by the Deutsche Forschungsgemeinschaft (DFG, German Research Foundation) under RTG 1995 and RTG 2247, within the Priority Program SPP 2244 ``2DMP'', under Germany's Excellence Strategy - Cluster of Excellence Matter and Light for Quantum Computing (ML4Q) EXC 2004/1 - 390534769 and - Cluster  of  Excellence and Advanced Imaging of Matter (AIM) EXC 2056 - 390715994. We acknowledge computational resources provided by the Max Planck Computing and Data Facility and RWTH Aachen University under project number rwth0811. This work was supported by the Max Planck-New York City Center for Nonequilibrium Quantum Phenomena. P.N. acknowledges support as a Moore Inventor Fellow through Grant No. GBMF8048 and gratefully acknowledges support from the Gordon and Betty Moore Foundation.
DNB is Moore Investigator in Quantum Materials EPIQS GBMF9455. DH was supported by a grant from the Simons Foundation (579913).

\section*{Author contributions}
D.H. conducted the SNOM experiments with supervision by D.N.B. S.T. conducted and analysed the STM and STS experiments with supervision by A.N.P. N.R.F. and V.S. fabricated the studied TDTG samples with supervision by J.Hone and C.D. J.Profe performed tight binding calculations with supervision by D.M.K. C.J.C. performed \emph{ab initio} calculations with supervision by P.N. D.H. developed the atomic relaxation code and performed related calculation and analysis as well as the nearfield tomography modelling in the Supplementary Information. K.W. and T.T. grew the hBN crystals. 
S.T. and D.H. wrote the manuscript with contributions from C.J.C., J.Profe, D.M.K and D.N.B. D.M.K and D.N.B supervised the project. All authors contributed to discussions and reviewed the manuscript.



\newpage
\begin{figure} [htbp]
\centering
\includegraphics[width=\linewidth]{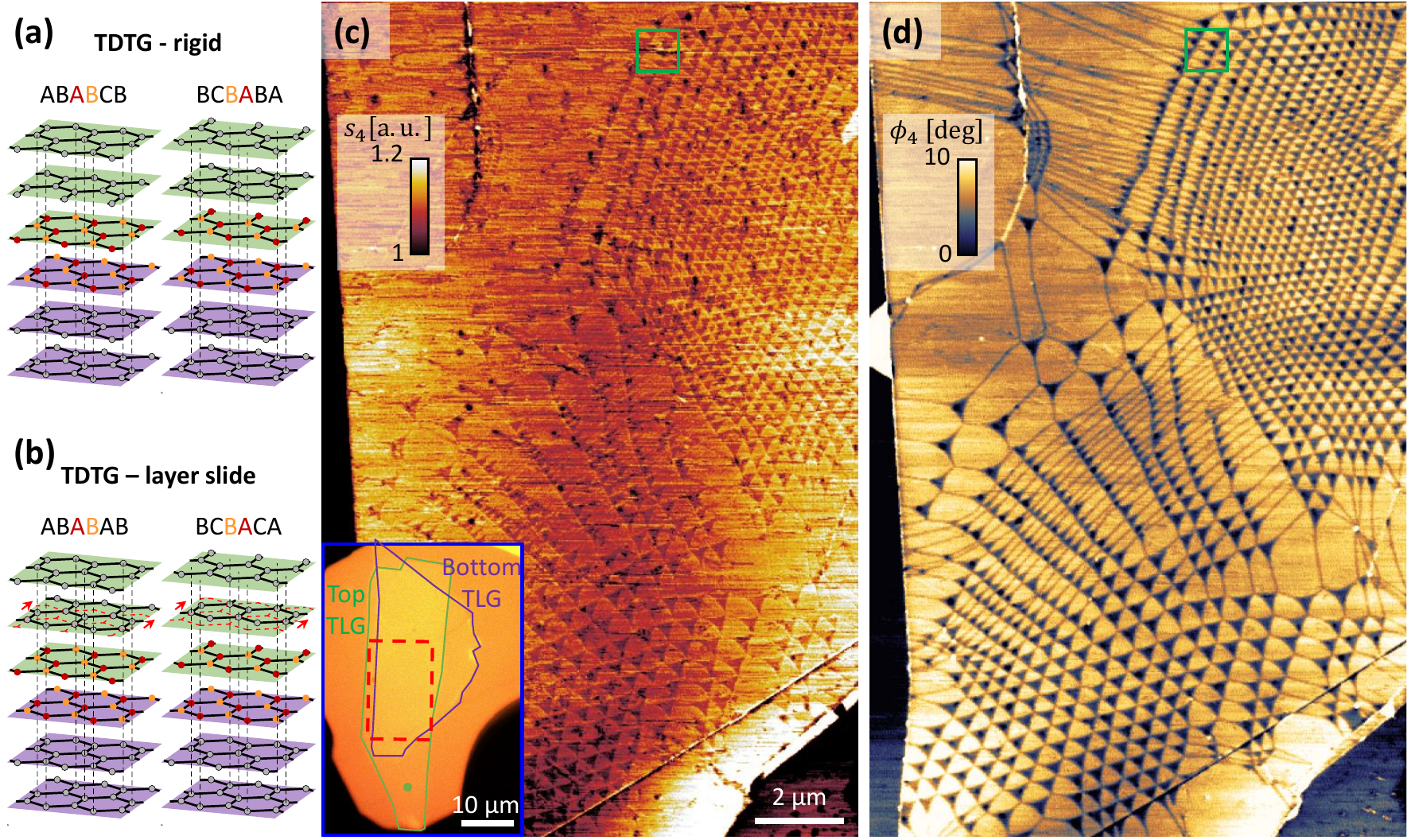}
\caption{\textbf{Moir\'e super-lattice of twisted double trilayer graphene (TDTG)}. \textbf{(a)} The lattice structure of the two lowest energy stacking configurations of TDTG when assuming both top and bottom trilayer graphene (TLG) are Bernal stacked. The interfacial atoms are colored to highlight the AB/BA stacking configurations. \textbf{(b)} The TDTG system after a global slide of the middle layer of one of the TLG sheets. The direction of slide is marked by red arrows indicating the transition from an ABA (dashed red honeycomb lattice) to a BAB configuration. Such a slide, if realized, results in the formation of the unstable BCBACA phase. \textbf{(c-d)} Mid-IR near-field amplitude (\textbf{c}) and phase (\textbf{d}) of the TDTG stack. Near-field signal in \textbf{c-d} is the 4th harmonic demodulation of the probe tapping frequency (laser illumination at $1000$~cm$^{-1}$, see Methods for additional experimental details). Inset of \textbf{c}: Optical image of the TDTG/hBN stack. The bottom and top TLGs are highlighted by purple and green frames respectively. The  red rectangle marks the scan area of \textbf{c-d}.}
\label{fig:TDTG_overview}
\end{figure}

\newpage
\begin{figure} [htbp]
\centering
\includegraphics[width=\linewidth]{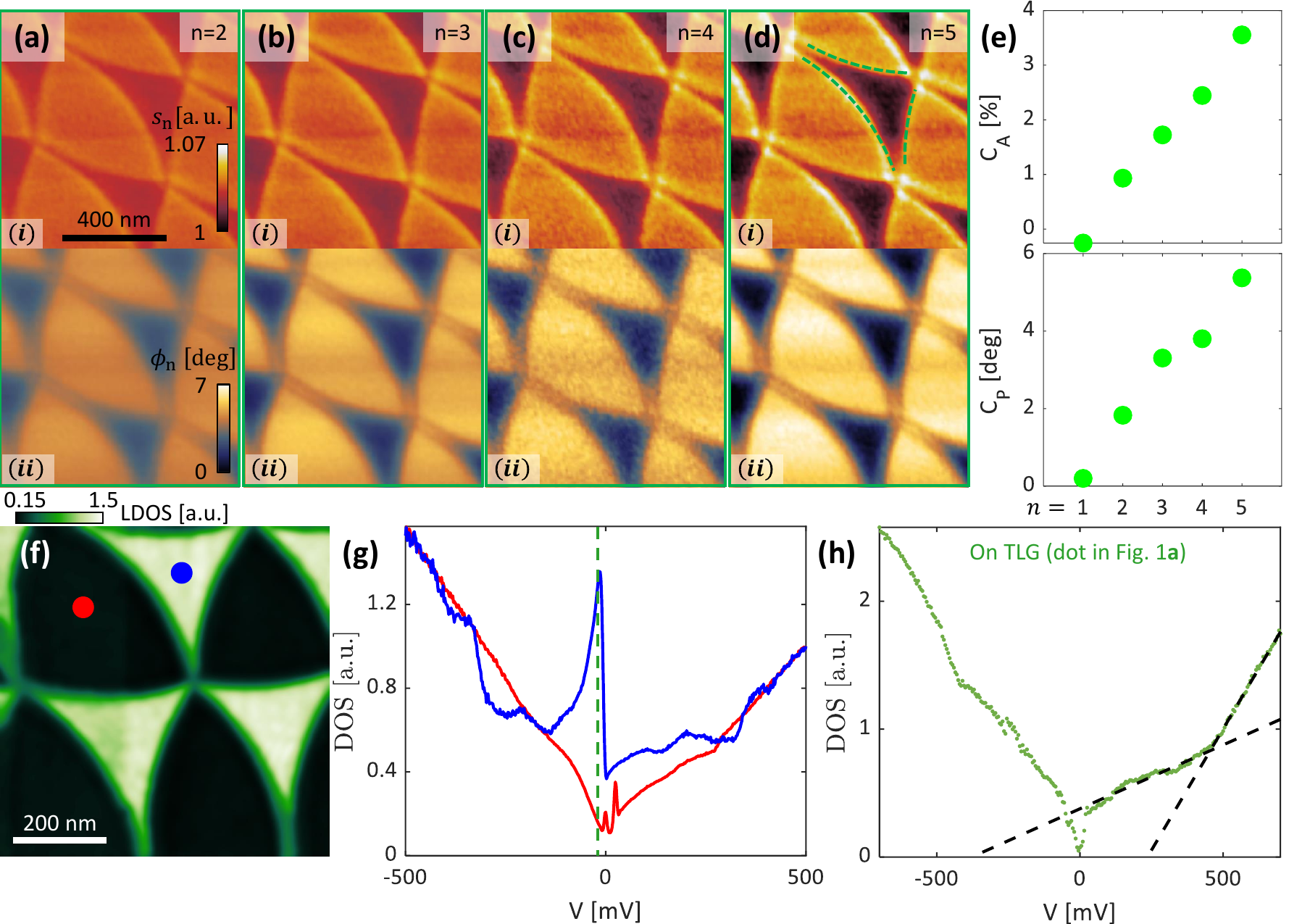}
\caption{\textbf{Imaging of the TDTG moir\'e}. \textbf{(a-d)} Mid-IR near-field imaging (over green square in Fig. \ref{fig:TDTG_overview}\textbf{a}): normalized amplitude $(i)$ and phase $(ii)$ at different probe demodulation harmonics from 2nd \textbf{(a)} to 5th \textbf{(d)}. The measurement was done at room temperature and a laser frequency of 1000~cm$^{-1}$. \textbf{(a-d)} share a color-bar and scale-bar. Dashed arcs in \textbf{(d)} with a radius of 850~nm highlight the domain shape. \textbf{(e)} Amplitude (top) and phase (bottom) contrasts between the two domains of \textbf{(a-d)} as a function of demodulation harmonics, revealing a monotonic trend. \textbf{(f)} Differential conductance measured by STS, acquired at a sample bias of $-20$~meV. \textbf{(g)} Tunneling spectrum measured in each of the two domains (marked by correspondingly colored dots in \textbf{(f)}). The dashed line marks the energy at which \textbf{(f)} was acquired. Each curve is normalized to its value at 500~mV. \textbf{(h)} Tunneling spectrum measured on exposed TLG section (green dot in Fig. \ref{fig:TDTG_overview}\textbf{a}) indicative of a Bernal stacked TLG source crystal.}
\label{fig:TDTG_tomography_experiment}
\end{figure}

\newpage

\newpage

\renewcommand{\figurename}{Supplementary Fig.}
\setcounter{figure}{0}

\section*{Supplementary Information for Unconventional Non-local Relaxation Dynamics in a Twisted Graphene Moir\'e Superlattice}

\bigbreak
\bigbreak

\section{Calculation of the optical conductivity}
To calculate the optical conductivity of graphene multilayers, we begin with the simplest possible tight-binding model Hamiltonian, which has proven to describe the optical properties of few-layer Graphene well~\cite{lui_imaging_2011, mak_evolution_2010}. This model consists of two hopping terms, the nearest-neighbor in-plane hopping $\gamma_0 = 3.16$\;eV and the nearest-neighbor out-of-plane hopping $\gamma_1 = 0.39$\;eV. All other hopping terms are neglected as they mainly alter the low energy region, which we are not accessing in the experiment. 
To consider the effect of electric fields on the measurement, we performed simulations in two different extreme settings. For the first case, we assumed that the outermost layers fully screen the applied field. This is incorporated into the Hamiltonian by a shift of $\pm \frac{e\phi}{2}$ of the chemical potential of the two outermost layers. For the second case, we assumed that each layer screens away the same amount of the field. This results in a decrease by $\frac{e\phi}{N}$ (where $N$ is the total number of layers: $N=6$ for TDTG) from layer to layer, such that the uppermost lies at $\frac{e\phi}{2}$- and the lowermost at $-\frac{e\phi}{2}$-potential. 

Once the tight-binding Hamiltonian is established we employ the standard Kubo formalism to calculate the  optical conductivity in the linear response regime~\cite{kubo_statistical-mechanical_1957}. In the following we will sketch the steps necessary to derive this formula. We use atomic units for convenience. The starting point for the derivation is a general non-interacting tight-binding Hamiltonian
\begin{equation}
    \hat{H} =  \sum_{\rm i,j,\sigma} t_{\rm i,j} c_{\rm i,\sigma}^{\dagger}c_{\rm j,\sigma} =\sum_{\rm m,n,\sigma, \bvec{k}} \;\sum_{\bvec{R},i,j} t_{\rm i,j} \delta_{r_{\rm m} +\bvec{R}, r_{\rm n} +\bvec{d_{\rm i,j}}}e^{-i\bvec{k}\cdot \bvec{R}}c_{\rm m,\sigma}^{\dagger}(\bvec{k})c_{\rm n,\sigma}(\bvec{k}),
\end{equation}
with $t_{\rm i,j}$ the tight binding hopping elements each associated with a vector $\bvec{d_{i,j}}$ connecting two lattice sites. The operators $c_{\rm i,\sigma}^{(\dagger )}$ annihilate (create) an electron on site $i$ with spin $\sigma$ and $c_{\rm m,\sigma}^{(\dagger )}(\bvec{k})$ annihilate (create) an electron on unit-cell site $m$ with spin $\sigma$ and momentum $\bvec{k}$. $\rm i,j$ denote positions in the full lattice and $\rm m,n$ denote positions within the unit-cell. The basis vectors spanning the real lattice are denoted as $\bvec{R}$. 
If we illuminate our sample with a weak laser we can incorporate it by performing a Peierls substitution
\begin{equation}
    t_{\rm i,j} \rightarrow t_{\rm i,j}e^{i\bvec{A}(t) \cdot \bvec{\delta}_{\rm i,j}}, 
\end{equation}
where the time dependent vector potential $\bvec{A}(t)$ has been introduced. The physical charge current generated by this perturbation can be extracted from a second order Taylor expansion in one of the three spatial directions $\alpha$ of the exponential and is given by 
\begin{align}
    \hat{\mathcal{J}}_{\alpha} = -\frac{\partial \hat{H}}{\partial A_{\alpha}(t)} = +\hat{j}_{\alpha}|_{\bvec{A}(t) = \bvec{0}}+A_{\alpha}(t)\hat{j}^D_{\alpha}|_{\bvec{A}(t) = \bvec{0}}.
\end{align}
Here, the diamagnetic current is defined as
\begin{equation}
    \hat{j}^D_{\alpha} = \sum_{\rm m,n,\sigma, \bvec{k}} \sum_{\bvec{R},i,j} (\bvec{d_{\rm i,j}})_{\alpha}^2\;t_{\rm i,j} \delta_{r_{\rm m} +\bvec{R}, r_{\rm n} +\bvec{\delta_{\rm i,j}}}e^{-i\bvec{k}\cdot \bvec{R}}c_{\rm m,\sigma}^{\dagger}(\bvec{k})c_{\rm n,\sigma}(\bvec{k})
\end{equation}
and the normal current as
\begin{equation}
    \hat{j}_{\alpha} = -i\sum_{\rm m,n,\sigma, \bvec{k}} \sum_{\bvec{R},i,j} (\bvec{d_{\rm i,j}})_{\alpha}\;t_{\rm i,j} \delta_{r_{\rm m} +\bvec{R}, r_{\rm n} +\bvec{\delta_{\rm i,j}}}e^{-i\bvec{k}\cdot \bvec{R}}c_{\rm m,\sigma}^{\dagger}(\bvec{k})c_{\rm n,\sigma}(\bvec{k}).
\end{equation}
From linear response theory we know that the optical conductivity is given by
\begin{equation}
    \sigma_{\alpha,\beta}(z) = \frac{i}{z} \braket{\hat{j}^D_{\alpha}} \delta_{\alpha,\beta}+ \frac{1}{zV}\int_0^{\infty} dt \, e^{izt}\braket{[\hat{j}_{\alpha}(t)_H,\hat{j}_{\beta}]}.
\end{equation}
The first term can be evaluated straight forward. For the second term, we assume that we know a diagonal basis of the Hamiltonian, whose eigenstates we label with $q,p$. This allows us to write 
\begin{align}
   \int_0^{\infty} dt \, e^{izt}\braket{[\hat{j}_{\alpha}(t)_H,\hat{j}_{\beta}]} &= 
    \sum_{\rm p,q} j^{pq}_{\alpha}j^{qp}_{\beta} \int_0^{\infty} dt \, e^{izt}e^{i(\epsilon_p-\epsilon_q)t} \braket{c_p^{\dagger} c_q c_q^{\dagger} c_p - c_p^{\dagger} c_q c_q^{\dagger} c_p} \\
    &=-i\sum_{\rm p,q} j^{pq}_{\alpha}j^{qp}_{\beta}\frac{n_f(\epsilon_p)-n_f(\epsilon_q)}{z+\epsilon_p-\epsilon_q}.
\end{align}
Collecting all contributions and performing an analytic continuation to $z = \omega + i \eta$ gives
\begin{align}
    \sigma_{\alpha,\beta}(\omega + i \eta) &= \frac{i}{\omega + i \eta} \sum_q j^D_{q,q}n_f(\epsilon_q) \delta_{\alpha,\beta}-i\sum_{\rm p,q} \frac{1}{(\omega + i \eta )V} j^{pq}_{\alpha}j^{qp}_{\beta}\frac{n_f(\epsilon_p)-n_f(\epsilon_q)}{(\omega + i \eta )+\epsilon_p-\epsilon_q}\\
    &= \frac{i}{\omega + i \eta} \sum_q j^D_{q,q}n_f(\epsilon_q) \delta_{\alpha,\beta}
        -\frac{i}{V}\sum_{\rm p,q}\frac{n_f(\epsilon_p)-n_f(\epsilon_q)}{\epsilon_p-\epsilon_q}j^{pq}_{\alpha}j^{qp}_{\beta}\left(\frac{1}{(\omega + i \eta )+\epsilon_p-\epsilon_q}-\frac{1}{(\omega + i \eta )}\right). \label{eq:full}
\end{align}
At low frequencies, the real part of this equation is governed by Lorentzian contributions, which can be associated to the Drude weight of the system. Relabeling the eigenstates $p,q$ by band and momentum indices we arrive at
\begin{align}
   \nonumber \sigma_{\alpha,\beta}(\omega) = &\frac{i}{\omega + i \eta} \sum_{b_1,\bvec k } j^D_{b_1,b_1}(\bvec k)n_f(\epsilon_{b_1}^{\bvec k}) \delta_{\alpha,\beta}\\
        &-\frac{i}{V}\sum_{b_1,b_2, \bvec k}\frac{n_f(\epsilon_{b_1}^{\bvec k})-n_f(\epsilon_{b_2}^{\bvec k})}{\epsilon_{b_1}^{\bvec k}-\epsilon_{b_1}^{\bvec k}}j^{b_1,b_2}_{\alpha}(\bvec k)j^{b_2,b_1}_{\beta}(\bvec k)\left(\frac{1}{(\omega + i \eta )+\epsilon_{b_1}^{\bvec k}-\epsilon_{b_1}^{\bvec k}}-\frac{1}{(\omega + i \eta )}\right). \label{eq:opt_cond}
\end{align}
As we have a layered structure we need to identify the optical sheet conductivity of each layer. For this purpose, we define the layered current operator according to
\begin{equation}
    j^{b_1,b_2}_{\alpha}(\bvec k)_{layer} = \frac{1}{2}\Braket{\epsilon_{b_1}^{\bvec k}|n^{\bvec k}}\left( j_{n,m}(\bvec k) \delta_{z_{layer},z_m} + j_{n,m}(\bvec k) \delta_{z_{layer},z_n} \right)\Braket{m^{\bvec k}|\epsilon_{b_2}^{\bvec k}}\label{eq:layer}
\end{equation}
Where we introduced two momentum-orbital-space unities $\ket{m^{\bvec k}}\bra{m^{\bvec k}}$. In this basis, $m$ labels the sites within the unit cell, therefore we can identify to which layer each site belongs and restrict the current operator to this specific layer. 
The layered conductivity $\tau_n$ is now obtained by introducing the definition~\eqref{eq:layer} in each term once:

\begin{align}
   \nonumber \tau_n^{\alpha,\beta}(\omega) = &\frac{i}{\omega + i \eta} \sum_{b_1,\bvec k } j^D_{b_1,b_1}(\bvec k)_{n}\;n_f(\epsilon_{b_1}^{\bvec k}) \delta_{\alpha,\beta}\\
        &-\frac{i}{V}\sum_{b_1,b_2, \bvec k}\frac{n_f(\epsilon_{b_1}^{\bvec k})-n_f(\epsilon_{b_2}^{\bvec k})}{\epsilon_{b_1}^{\bvec k}-\epsilon_{b_1}^{\bvec k}}j^{b_1,b_2}_{\alpha}(\bvec k)_{n}\;j^{b_2,b_1}_{\beta}(\bvec k)\left(\frac{1}{(\omega + i \eta )+\epsilon_{b_1}^{\bvec k}-\epsilon_{b_1}^{\bvec k}}-\frac{1}{(\omega + i \eta )}\right). \label{eq:layer_opt_cond}
\end{align}

These layer conductivities (defined in eq. \ref{eq:layer_opt_cond}) sum to the full conductivity. We chose $T = 0.025$\,eV and a phenomenological broadening $\eta = 30$\,meV. In practice we represent the conductivity in units of $\sigma_0 = \frac{\pi e^2}{2h}$, the optical conductivity of a single layer graphene. 

\section{Evidence the source crystal is Bernal trilayer graphene}

\begin{figure} [htbp]
\centering
\includegraphics[width=4.4in]{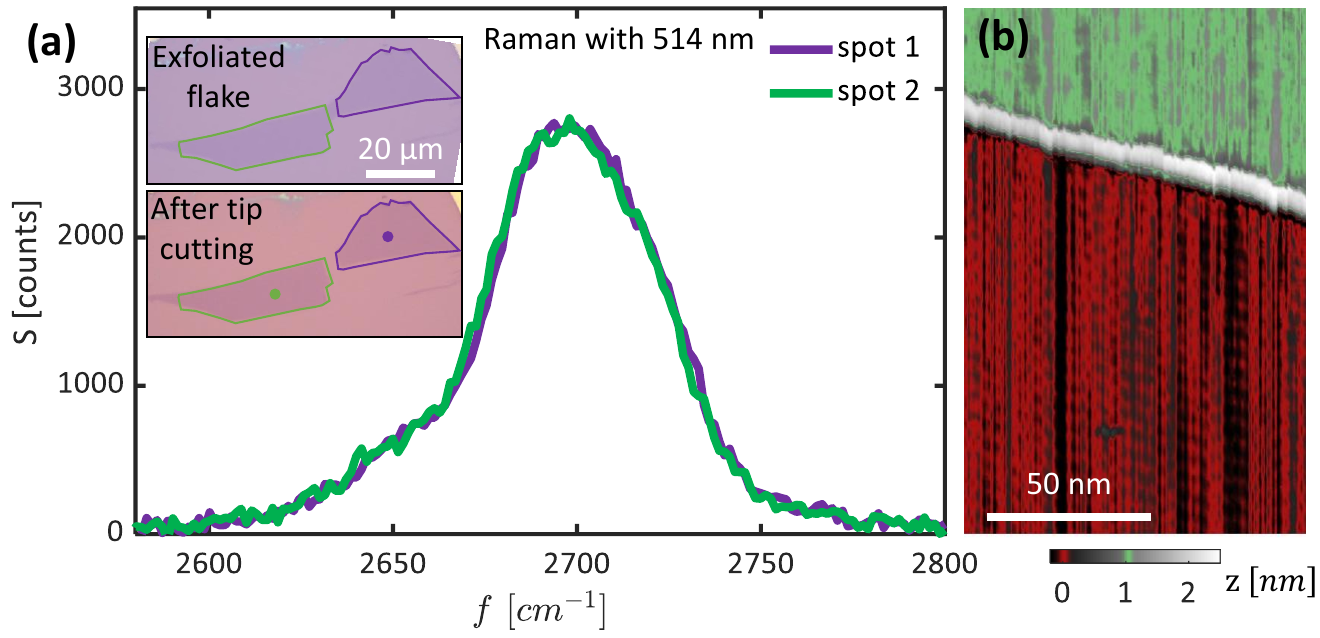}
\caption{\textbf{Evidence the source crystal is Bernal trilayer graphene}. \textbf{(a)} Raman spectra of the 2D-mode acquired on two different spots of the source crystal with 514 nm laser wavelength.  Inset: optical micrographs of the source crystal before (top) and after (bottom) AFM cutting.  Blue and green spots indicate where Raman spectra were obtained.  \textbf{(b)} STM topograph of the fabricated TDTG stack showing a trilayer step at the edge of the twisted region.}
\label{fig:TLG_bernal_proof_ver2}
\end{figure}

Prior to device fabrication, the thickness of the source graphene crystal was determined using its optical contrast against the SiO$_2$ exfoliation substrate in the standard way (Supplementary Fig. \ref{fig:TLG_bernal_proof_ver2}a, inset).  We checked that the source crystal was Bernal rather than rhombohedral by performing Raman spectroscopy on several areas.  The Raman 2D-mode of few layer graphene exhibits distinctive features depending on the stacking configuration, with rhombohedral graphene showing a pronounced asymmetric line shape \cite{lui_imaging_2011,Cong2011}.  In our Raman measurements we observe no features indicative of rhombohedral graphene, confirming that our source crystal was stacked in the Bernal configuration (Supplementary Fig. \ref{fig:TLG_bernal_proof_ver2}a).  We further confirmed the thickness of the source crystal as trilayer by measuring STM topography across the step leading to the twisted region of the device (Supplementary Fig. \ref{fig:TLG_bernal_proof_ver2}b).  Spectroscopy measured below the step provides additional evidence that the source crystal was Bernal stacked, as described in the main text (Fig. \ref{fig:TDTG_tomography_experiment}h).

\section{Alternative mechanisms for the moir\'e contrast and feasibility of atomic-scale near-field tomography}
\label{nearfield_tomography}

The observation of strong contrasts in the optical conductivity and the local density of states, and the difference in structural stability between the moir\'e domains in TDTG implies either that the stacking configurations in our device follow the rigid scenario (Fig. \ref{fig:TDTG_overview}a) and that an additional $C_2^z$ symmetry breaking mechanism is present, or that the layer slide scenario (Fig. \ref{fig:TDTG_overview}b) is realized through a non-local relaxation effect.  We have considered several possible $C_2^z$ symmetry breaking mechanisms in an attempt to explain the observed moir\'e contrast.  One source of $C_2^z$ symmetry breaking could have been an out-of-plane displacement field caused by trapped charge in the substrate.  We can rule this out because such substrate inhomogeneities would likewise dope the graphene away from charge neutrality, which would in turn be clearly visible in STS measurements as an energy shift of the charge neutrality point.  In our measured spectrum on exposed trilayer (Fig. \ref{fig:TDTG_tomography_experiment}h), however, the minimum in the density of states is pinned to zero energy, indicating negligible intrinsic doping and hence negligible displacement field.  Another potential source of $C_2^z$ symmetry breaking is the asymmetric dielectric environment imposed by the substrate itself, since the presence of hBN on one side of the device and vacuum on the opposite side does in principle break $C_2^z$ symmetry.  To investigate whether this asymmetry can lead to significant differences in the electronic properties or stacking energies of the two domains we have performed first principles Density Functional Theory (DFT) calculations (see Methods for details) and found that the hBN substrate introduces an energy imbalance of at most $70~\mu\textrm{eV/nm}^2$, which would produce a DW radius of curvature of over $40~\mu\textrm{m}$.  The true radius of curvature as measured in SNOM and STM is $\sim850$ nm (see overlaid arcs in Fig. \ref{fig:TDTG_tomography_experiment}d), an order of magnitude less than that predicted due to the substrate effect, ruling out the latter as an explanation of our data.\\ 

The only remaining plausible $C_2^z$ symmetry breaking mechanism in our experiments is in the geometry of the scanned probes themselves.  In both SNOM and STM, the signal is collected from a metallic probe that is suspended above the sample.  As a result, measurements of multilayer samples tend to be most sensitive to the properties of those layers that are physically closest to the probe.  In STM, this can lead to exponential sensitivity to the density of states of the topmost layer due to the rapid decay of the surface wave function across the vacuum tunneling barrier \cite{Kerelsky2021,Rubio-Verd2022}.  The SNOM signal, on the other hand, is generated by a classical electric field that follows a power law decay, which makes it possible to perform tomographic imaging of bulk materials by controlling the height of the probe \cite{Sun2009,Govyadinov2014b,Mooshammer2018}.  This asymmetric probe sensitivity has been exploited to attain imaging resolution along the vertical ($z$) direction, using a technique called near-field tomography (NFT).  NFT utilizes the fact that the $z$-profile of the complex optical conductivity is imprinted in the near-field probe approach curve\cite{Taubner2005,Sun2009}. NFT was recently used, for example, to decouple surface and bulk contributions to the near-field signal in a topological insulator\cite{Mooshammer2018}.\\ 

\begin{figure} [htbp]
\centering
\includegraphics[width=6.05in]{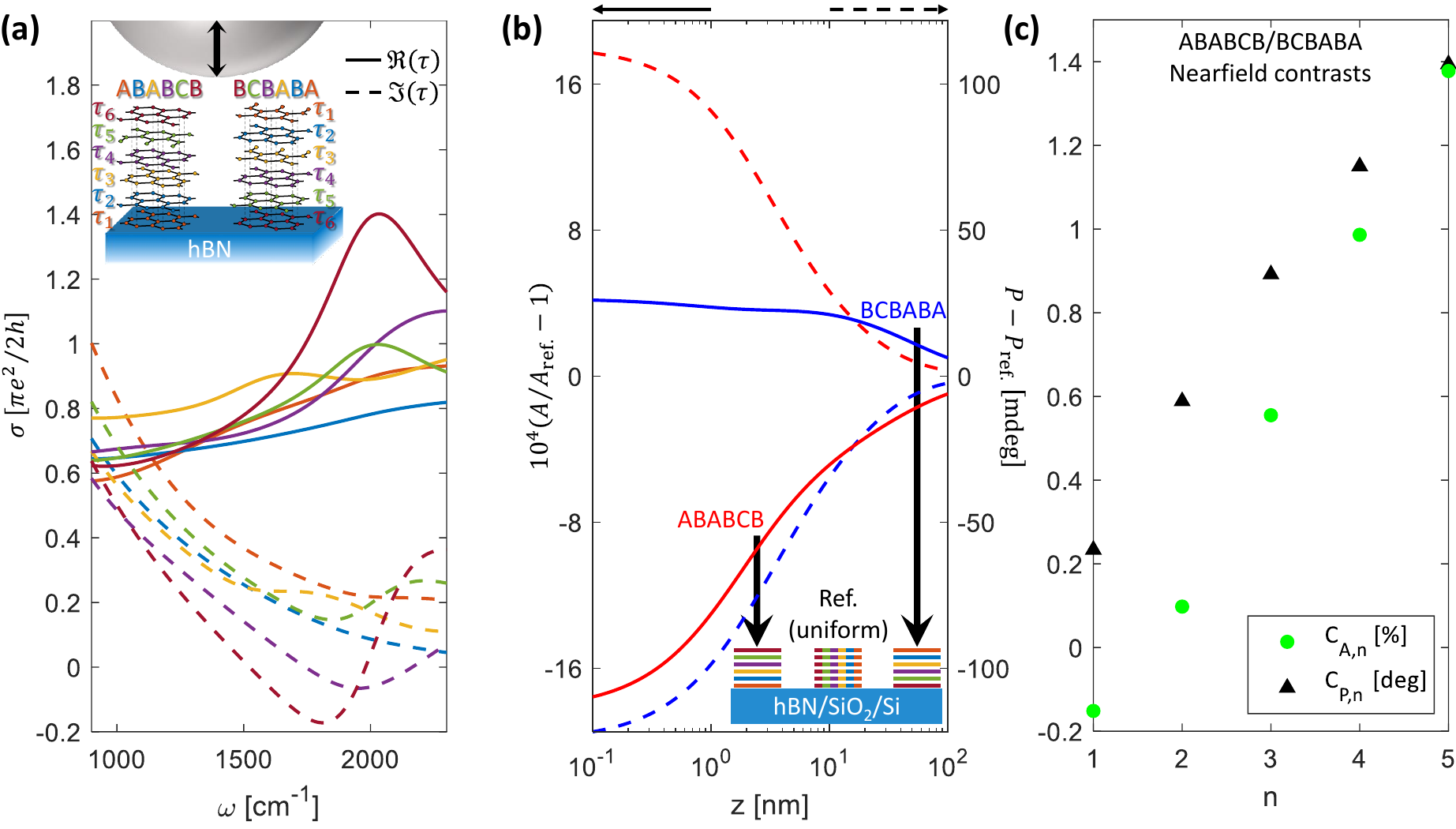}
\caption{\textbf{Atomic scale near-field tomography in ABABCB/BCBABA TDTG.} \textbf{(a)} Real (solid) and imaginary (dashed) parts of the layer resolved complex optical conductivity ($\tau_i$) for each of the six layers of ABABCB stacked graphene.  Layer conductivity assumes $E_F=60~\textrm{meV}$ and a $5~\textrm{meV}$ broadening (see Supplementary Information section 1 for additional details). Inset: schematic of near-field tomography showing that the enhanced localized field under the tapping probe (not to scale) breaks $C_2^z$ symmetry and generates a contrast in the near-field response. \textbf{(b)} Comparing amplitude (left, solid) and phase (right, dashed) of approach curves. The ABABCB (red) and the BCBABA (blue) approach curves are presented with respect to a reference stack with similar total conductivity but distributed uniformly across the layers (see Supplementary Information section 1 for calculation details). Inset: schematics of the three discussed configurations; colors match those in \textbf{(a)}. \textbf{(c)} The ABABCB/BCBABA contrast as a function of tapping frequency harmonics as derived from the approach curves of \textbf{(b)} for a similar tapping amplitude as in Fig. \ref{fig:TDTG_tomography_experiment}.}
\label{fig:TDTG_tomography_theory}
\end{figure}

Until now it has been assumed that the vertical resolution of NFT is limited by the curvature of the AFM tip, which is typically several tens of nanometers.  This limitation arises from the radius of curvature setting the decay length of the electromagnetic (EM) field away from the tip.  If, however, a device architecture can be engineered so as to reduce or eliminate far field contrast, then the resolution of NFT can be significantly enhanced.  In fact, we find that a heterostructure consisting of ABABCB and BCBABA domains consitutes just such a system, in which the vertical resolution of NFT can be boosted to sub-nanometer length scales.  We define a conductivity $\tau_i$ for layer $i=1,...,6$ (Supplementary Fig. \ref{fig:TDTG_tomography_theory}a), consistent with the Hamiltonian of the system (Supplementary Information section 1). From this definition the ABABCB and BCBABA phases are modelled by stacking $\tau_i$ with the appropriate layer order, separated by the layer spacing of graphite (Inset of Supplementary Fig. \ref{fig:TDTG_tomography_theory}a). The observed near-field contrast is uniquely generated by an EM $z$-gradient and will not have any contribution from a uniform EM field, strictly due to the $C_2^z$ symmetry relation between ABABCB and BCBABA stackings. Plugging the optical conductivities of Supplementary Fig. \ref{fig:TDTG_tomography_theory}a into the lightning rod model solver\cite{PhysRevB.90.085136} (LRM) we get small differences between the ABABCB and BCBABA approach curves (Supplementary Fig. \ref{fig:TDTG_tomography_theory}b). After demodulation, these approach curves result in a near-field contrast for different probe harmonics (Supplementary Fig. \ref{fig:TDTG_tomography_theory}c) well within our measurement capabilities (Fig. \ref{fig:TDTG_tomography_experiment}e).  We emphasize that the ability to resolve the atomic scale tomographic differences between ABABCB and BCBABA stacking configurations is a direct consequence of the $C_2^z$ symmetry relation between these domains, which eliminates all far field optical contrast.  While we can rule out NFT as a description of our experimental results, as described in the main text, the foregoing demonstrates that NFT at the atomic scale is in principle achievable in device geometries with mirror symmetric designs.

\section{Possible TDTG configurations where the moir\'e superlattice domains are C2 symmetry pairs}
\label{c2_pairs}

Here, we consider the eight stacking configurations for TDTG that were omitted in the main text.  These eight configurations, displayed in Supplementary Fig. \ref{fig:TDTG_remaining_configurations}, constitute four moir\'e pairs that are related by $C_2^z$ symmetry.  Unsurprisingly, these moir\'e pairs exhibit no contrast in stacking energy or in electronic structure, and therefore cannot be accurate descriptions of the stacking configurations realized in our experimental device.

\section{The optical contrast of the moir\'e superlattice in TDBG}

Several prior experimental works have reported near-field optical contrast between domains of Bernal (ABAB) and rhombohedral (ABCA) few layer graphene\cite{Halbertal2021,Kerelsky2021}, similar to the contrast between ABABAB and BCBACA domains observed in the present work (Fig. \ref{fig:TDTG_overview}).  This observed contrast is typically attributed to differences in the optical conductivity between Bernal and rhombohedral graphenes.  Such an attribution is plausible, given the significant difference in electronic structure between the two stacking configurations, however it has not yet been justified from a theoretical perspective.  Here, we demonstrate that the observed near-field contrast in domains of marginally twisted double bilayer graphene is quantitatively consistent with the calculated optical conductivities of the ABAB and ABCA stacking configurations.\\

\begin{figure} [htbp]
\centering
\includegraphics[width=4.95in]{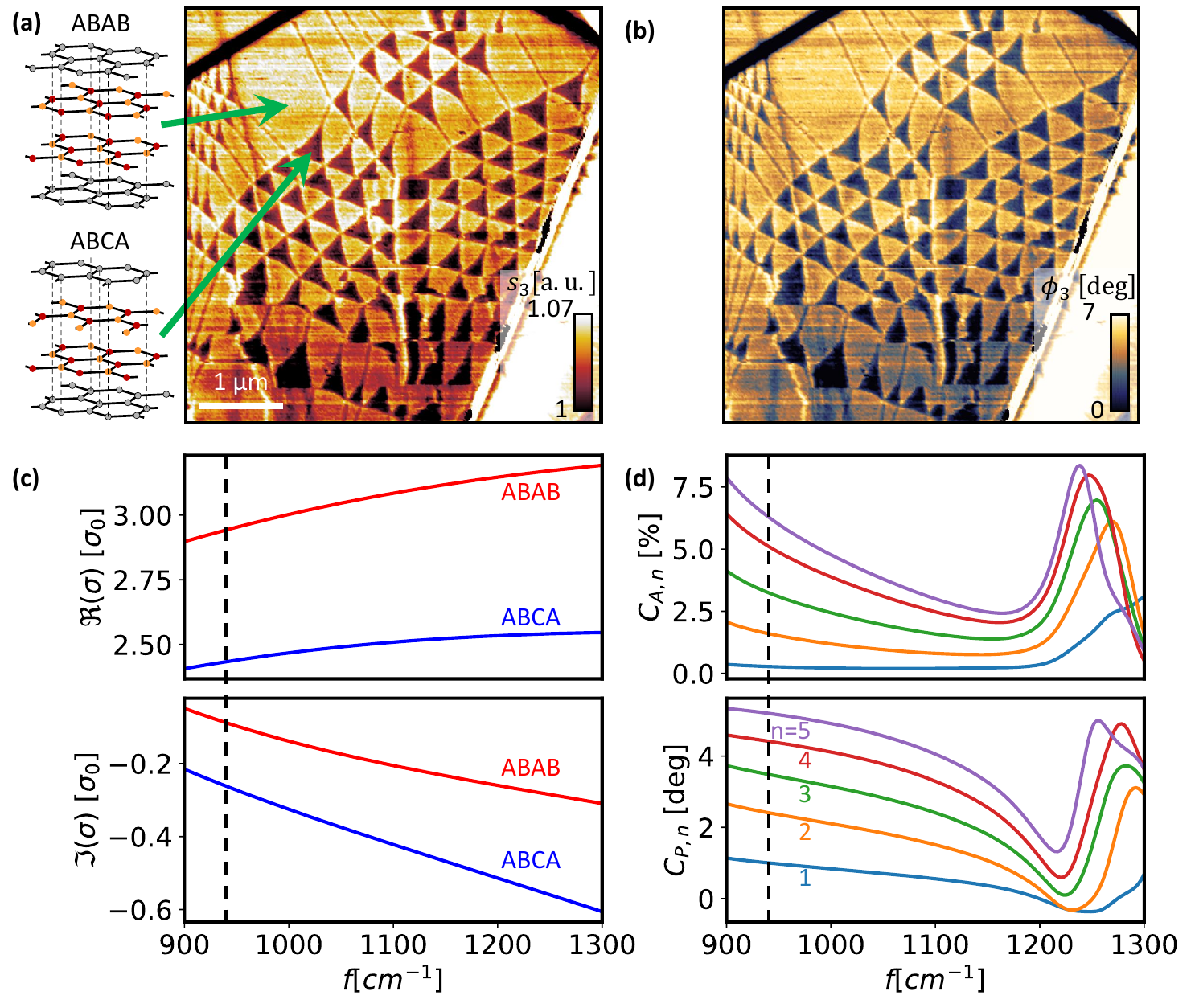}
\caption{\textbf{Modelling SNOM contrast in TDBG}. \textbf{(a} and \textbf{b)} Near-field amplitude and phase scans of a marginally twisted TDBG sample, acquired at $940~\textrm{cm}^{-1}$ in the third tapping harmonic.  Triangular domains correspond to ABAB (light) and ABCA (dark) stacking configurations, as illustrated on the left. \textbf{(c)} Real (top) and imaginary (bottom) parts of the frequency dependent optical conductivities of ABAB (red) and ABCA (blue) graphene derived from tight binding calculations of the band structures, plotted in units of $\sigma_0=\frac{\pi e^2}{2h}$.  \textbf{(d)} LRM calculations of the frequency dependent near-field contrast between ABAB and ABCA domains for the lowest five tapping harmonics, $n$.  Amplitude and phase contrasts are shown in top and bottom plots, respectively.  Vertical dashed line indicates the frequency at which (a) and (b) were acquired.  LRM calculations assume a tip with $30~\textrm{nm}$ radius of curvature and $40~\textrm{nm}$ tapping amplitude above the surface of ABAB or ABCA graphene on hBN ($40~\textrm{nm}$ thick) on SiO$_2$ ($285~\textrm{nm}$) on a Si substrate.}
\label{fig:TDBG_figure}
\end{figure}

Supplementary Fig. \ref{fig:TDBG_figure}a,b illustrates the contrast in amplitude and phase observed in SNOM measurements of TDBG.  In these images, the darker (brighter) triangles correspond to ABCA (ABAB) stacking, as evidenced by the concavity (convexity) of their edges.  In order to explain this contrast, we use the tight binding derived band structures of the ABAB and ABCA stacking configurations (see Supplementary Information section 1) to compute the frequency dependent complex optical conductivity of each domain, shown in Supplementary Fig. \ref{fig:TDBG_figure}c.  We then simulate the near-field signal with the lightning rod model (LRM) \cite{PhysRevB.90.085136}, using a tip radius of 30~nm and tapping amplitude of 40~nm for each stacking configuration.  The LRM calculation assumes that each stacking configuration is placed on a 40~nm thick hBN substrate on top of 285~nm thick SiO$_2$ on Si.  This results in a frequency dependent amplitude and phase for each domain.  To facilitate comparison with experiment, we examine the amplitude and phase contrast between ABCA and ABAB domains within the LRM calculation, shown in Supplementary Fig \ref{fig:TDBG_figure}d for the lowest five tapping harmonics.  The frequency at which the experimental images in Supplementary Fig. \ref{fig:TDBG_figure}a,b were acquired is marked with a vertical dashed line.  The calculated contrast in both the amplitude and phase channels shows good quantitative agreement with the measured contrast, demonstrating that the contrast observed in the experimental images can be generated by the different band structures and thus optical conductivities of the two stacking configurations.

\newpage
\begin{table}[ht]
\centering
\caption{Stacking energy for the two lowest energy configurations for different moir\'e systems (in units of $\frac{\mathrm{meV}}{\mathrm{nm}^2}$).}
\label{tab:E_lowest_conf}
\vspace{0.5em}
\def\arraystretch{1.5}
\begin{tabular}{|c|c| c | c|}
    \hline
    \bfseries ABABCB &\, 5.059 &\, \bfseries BCBABA  &\, 5.059 \\\hline
    \bfseries ABABCA &\, 9.755 &\, \bfseries BCBABC  &\, 9.883 \\\hline
    \bfseries BCBACB &\, 14.746 &\, \bfseries ABABAC  &\, 4.768 \\\hline
    \bfseries BCBACA &\, 10.042 &\, \bfseries ABABAB  &\, 0 \\\hline
    \bfseries BCABCA &\, 19.420 &\, \bfseries CABABC  &\, 9.553 \\\hline
\end{tabular}
\end{table}


\newpage
\begin{table}[ht]
\centering
\caption{GSFE parameters ($c_{0,\dots,5}$ - following nomenclature of Ref. \citeonline{carr:2018:relaxation} and units of $\frac{\mathrm{meV}}{\mathrm{nm}^2}$) for different moir\'e configurations considered in this work.}
\label{tab:material_param}
\vspace{0.5em}
\def\arraystretch{1.5}
\begin{tabular}{c|cccccc}
    & $c_0 \, $ & $c_1 \, $ & $c_2 \, $ & $c_3 \, $ & $c_4 \, $ & $c_5 \, $ \\\hline
    \bfseries ABABCB/BCBABA &\, 141.38 &\, 81.94 &\, -7.55 &\, -2.79 &\, 0 &\, 0 \\\hline
    \bfseries ABABCA/BCBABC &\, 142.14 &\, 80.61 &\, -5.62 &\, -3.65 &\, -0.0152 &\, -0.0095 \\\hline
    \bfseries BCBACB/ABABAC &\, 142.13 &\, 80.62 &\, -5.64 &\, -3.64 &\, 1.179 &\, 0.742 \\\hline
    \bfseries BCBACA/ABABAB &\, 141.76 &\, 81.26 &\, -6.57 &\, -3.23 &\, 1.186 &\, 0.747 \\\hline
    \bfseries BCABCA/CABABC &\, 142.50 &\, 79.98 &\, -4.71 &\, -4.05 &\, 1.166 &\, 0.734 \\\hline
\end{tabular}
\end{table}

\newpage

\newpage

\newpage
\begin{figure} [htbp]
\centering
\includegraphics[width=3.33in]{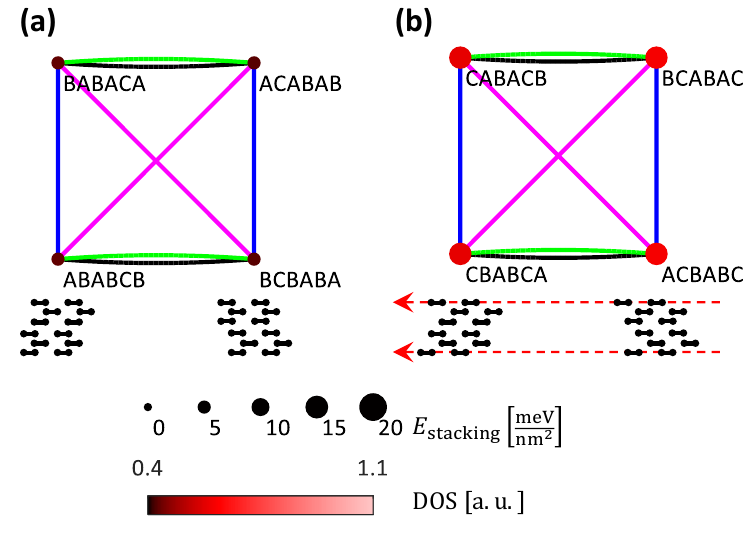}
\caption{\textbf{Exploration of additional TDTG moir\'e superlattice structures}. \textbf{(a-b)} Each panel addresses a group of TDTG configurations which were not discussed in the main text. Each moir\'e-pair is connected by a horizontal black line. Configurations that are $C_2$ symmetry pairs are connected by blue ($C_2^x$), green ($C_2^z$) and magenta ($C_2^xC_2^z$) lines. Each configuration is marked by a circle whose color indicates Fermi level spectral weight, and whose size indicates the configuration's stacking energy density (see legend).  Inset - bottom: Schematics of the configurations. The red arrow indicates the required global layer sliding in order to realize the particular moir\'e superlattice from the nominal ABABCB/BCBABA configuration.}
\label{fig:TDTG_remaining_configurations}
\end{figure}

\newpage

\end{document}